\documentclass[twocolumn, times, astrosymb, twocolappendix]{aastex631}
\usepackage{bm}
\usepackage{amsmath}
\usepackage{amssymb}
\usepackage{comment}
\usepackage{enumitem}
\usepackage{hyperref}
\usepackage{pifont}
\newcommand{\cmark}{\ding{51}}   
\newcommand{\xmark}{\ding{55}}   
\newcommand{\tmark}{\ding{115}}  

\revised{\today}
\accepted{\today}
\submitjournal{ApJ}

\begin{document}
\title{\texttt{microJAX}: A Differentiable Framework for Microlensing Modeling with GPU-Accelerated Image-Centered Ray Shooting}

\correspondingauthor{Shota Miyazaki}
\email{miyazaki@ir.isas.jaxa.jp}

\author[0000-0001-9818-1513]{Shota Miyazaki}
\affiliation{Institute of Space and Astronautical Science, Japan Aerospace Exploration Agency, 3-1-1 Yoshinodai, Chuo, Sagamihara, Kanagawa 252-5210, Japan}

\author[0000-0003-3309-9134]{Hajime Kawahara}
\affiliation{Institute of Space and Astronautical Science, Japan Aerospace Exploration Agency, 3-1-1 Yoshinodai, Chuo, Sagamihara, Kanagawa 252-5210, Japan}

\shorttitle{microJAX: differentiable microlensing}
\shortauthors{Miyazaki \& Kawahara}

\begin{abstract}
We introduce \texttt{microJAX}, the first fully differentiable implementation of the image-centered ray-shooting (ICRS) algorithm for gravitational microlensing. Built on \texttt{JAX} and its XLA just-in-time compiler, \texttt{microJAX} exploits GPU parallelism while providing exact gradients through automatic differentiation. The current release supports binary- and triple-lens geometries, including limb-darkened extended-source effects, and delivers magnifications that remain differentiable for all model parameters. Benchmarks show that \texttt{microJAX} matches the accuracy of established packages and attains up to a factor of $\sim$5-6 speed-up in the small-source, limb-darkened regime on an NVIDIA A100 GPU. Since the model is fully differentiable, it integrates seamlessly with probabilistic programming frameworks, enabling scalable Hamiltonian Monte Carlo and variational inference workflows. Although the present work focuses on standard microlensing magnification models, the modular architecture is designed to support upcoming implementations of microlensing higher-order effects, while remaining compatible with external likelihood frameworks that incorporate advanced noise models. \texttt{microJAX} thus provides a robust foundation for precise and large-scale surveys anticipated in the coming decade, including the \textit{Nancy Grace Roman Space Telescope}, where scalable, physically self-consistent inference will be essential for maximizing scientific return.
\end{abstract}


\section{Introduction \label{sec:intro}}
Recent advances in differentiable programming and probabilistic inference have opened new possibilities for precision modeling in astronomy. Tools such as \texttt{Theano} \citep{Theano2016}, \texttt{Tensorflow} \citep{tensorflow2015-whitepaper}, \texttt{JAX} \citep{jax2018github}, and \texttt{PyTorch} \citep{Paszke+2019} allow the construction of fully differentiable computational graphs, supporting high-performance automatic differentiation. When combined with probabilistic programming languages (PPLs) such as \texttt{Stan} \citep{Stan2017} and \texttt{NumPyro} \citep{Bingham+2018, Phan+2019}, these frameworks enable the direct application of efficient gradient-based optimization and inference methods, such as Hamiltonian Monte Carlo \citep[HMC;][]{Duane+1987, Hoffman+2011, Betancourt2017}, to complex scientific models. This has led to successful applications with high-dimensional Bayesian inference, for example, in exoplanetary system modeling \citep{Agol+2021, Masuda+2024}, X-ray spectral fitting \citep{Dupourqu+2024}, atmospheric retrievals of exoplanets \citep{Kawahara+2022, Kawahara+2025}, gravitational waves \citep{Isi+2021, Wong+2023, Edwards+2024}, and so on.\footnote{ \url{https://github.com/JAXtronomy/awesome-JAXtronomy}}\footnote{\url{https://juliaastro.org/home/}} With the diversification of scientific goals and the advent of next-generation surveys, microlensing modeling faces new demands. The upcoming \textit{Nancy Grace Roman Space Telescope} \citep{Spergel+2015, Akeson+2019} will provide high-cadence, high-precision light curves for tens of thousands of microlensing events, including planetary systems, dark lenses, and multi-component configurations \citep{Bennett+2002, Penny+2019, Lam+2023}. To fully leverage this dataset, modeling frameworks must handle diverse lens geometries, extended-source effects, complex surface brightness profiles, and other higher-order effects while remaining scalable and compatible with gradient-based inference. These challenges motivate the development of robust numerical methods that integrate differentiable programming techniques.

Modeling microlensing magnification becomes significantly more challenging when both multiple-lens geometries and finite-source effects are involved. While point-source calculations can be handled analytically or with efficient root-finding methods, extended-source models require numerical integration over the lensed images. There are two main approaches commonly used: contour integration \citep{Gould+1997, Dominik1998} and image-centered ray shooting \citep{Bennett+1996, Dong+2006, Bennett+2010}. The contour integrating method utilizes Green's theorem to transform the two-dimensional image integral into a one-dimensional integral over the boundaries of the image. This technique is highly efficient and has been widely adopted in binary-lens modeling, particularly in the well-established public package \texttt{VBBinaryLensing} \citep{Bozza2010, Bozza+2018}. However, although great efforts have been made to extend the contour integration technique to more complex lens configurations, along with high accuracy and performance \citep{Bozza+2025}, the method inherently encounters difficulties in determining the correct orientation and connectivity of image boundaries as the number of lenses increases, which is essential for applying Green's theorem. Moreover, for sources with spatially complex brightness profiles, the integrals require finer quadrature and significantly higher computational cost \citep{Dominik1998}.

In contrast, the image-centered ray shooting method \citep[ICRS;][]{Bennett+1996, Bennett+2010} straightforwardly performs numerical integration directly on the image plane by mapping rays back to the source plane. This approach is more flexible in handling complex lens configurations and is naturally suited for arbitrary surface brightness profiles, such as those with starspots \citep[e.g.,][]{Hendry+2002, Giordano+2015}. Its robustness near caustics and general applicability make it a powerful alternative to contour integration. However, the method typically requires dense sampling over each image, thus posing higher computational costs, particularly for small sources where the magnification, defined as the image area per unit source area, demands finer resolution of image areas to maintain accuracy. On the other hand, the inverse ray shooting \citep{Kayser+1986, Wambsganss1997} is inherently parallelizable across image-plane pixels; a variety of GPU-accelerated schemes have been developed for cosmological microlensing \citep[e.g., ][]{Thompson+2014, Zheng+2022, Weisenbach2025}. However, only a few exploratory GPU ports of the ICRS algorithm have been published \citep{Hundertmark2011, McDougall2016}, and none has yet undertaken a systematic optimisation of accuracy versus computational performance, which likely reflects the difficulty of adaptive image-centered mapping onto GPU architectures.

To address these challenges, we present \texttt{microJAX} \citep{microjax011}, a differentiable, GPU-accelerated microlensing modeling framework based on the ICRS method. The key contributions of this work are summarized as follows:
\begin{itemize}[itemsep=0.0ex, topsep=0.0ex, leftmargin=*]
    \item \textbf{Differentiable ICRS on GPU}:
    We implement the first fully differentiable ICRS algorithm compatible with JAX, enabling efficient automatic differentiation. The code leverages XLA compilation to exploit GPU parallelism while maintaining a static computation graph, which facilitates fast and stable magnification evaluations.
    \item \textbf{Fast and High-precision modeling of complex lens systems}:
    \texttt{microJAX} supports binary and triple lens systems with arbitrary surface brightness profiles. Using a polar grid discretization and smooth interpolation near image boundaries, we achieve relative magnification errors $\lesssim 10^{-4}$ even for small sources.\footnote{This accuracy refers to the default configuration and is not a fundamental limit. Higher precision can, in principle, be obtained by increasing the radial and angular resolutions in the ray shooting at the expense of computation time and memory usage.}
    \item \textbf{End-to-end gradient-based inference}:
    The framework is designed to interface seamlessly with probabilistic programming tools. We demonstrate HMC inferences on both synthetic and real microlensing light curves, showing that \texttt{microJAX} can serve as a building block for full Bayesian pipelines.
\end{itemize}
In the context of prior work, our approach differs from two dominant lines of development. Traditional ICRS implementations (e.g., \citealt{Bennett+1996, Bennett+2010}) are not differentiable and have not been optimized for GPU acceleration. Conversely, recent JAX-based codes such as \texttt{caustics} \citep{Bartolic2023} and \texttt{microlux} \citep{Ren+2025} achieve differentiability by adapting contour integration techniques. Still, they are currently limited to binary lenses and/or rely on adaptive sampling structures, which pose challenges for parallelization. \texttt{microJAX} thus offers a unified solution that is both scalable and differentiable, bridging these two approaches. \texttt{microJAX} is available on GitHub as a Python library with documentation and GPU support, aiming to provide a foundation for further development and community contributions.\footnote{\url{https://github.com/ShotaMiyazaki94/microjax}}

In this paper, we describe the design and implementation of \texttt{microJAX}, benchmark its computational performance and accuracy against existing tools, and demonstrate its applicability in both idealized and realistic inference scenarios. Section~\ref{sec:background} reviews the fundamentals of microlensing magnification modeling and introduces the image-centered ray shooting (ICRS) method. Section~\ref{sec:implementation} details the differentiable implementation of \texttt{microJAX} using \texttt{JAX}. Section~\ref{sec:benchmark} presents computational benchmarks and accuracy comparisons with existing microlensing tools on both CPU and GPU architectures. In Section~\ref{sec:application}, we illustrate the integration of \texttt{microJAX} into gradient-based inference workflows using synthetic light curves. Finally, Section~\ref{sec:conclusion} summarizes the key results and discusses future directions.

\section{Background \label{sec:background}}
To motivate the design of our modeling framework, we first summarize the mathematical foundations of gravitational microlensing with multiple lenses. In particular, we review the lens equation and its implications for the image formation process, which forms the basis for our inverse-ray shooting approach. We then briefly describe the image-centered ray shooting (ICRS) method, which serves as the computational backbone of \texttt{microJAX}.

\subsection{Multiple Lens Microlensing}
Assuming that all lenses are point masses, the lens equation for an $N$-body system can be written \citep{Kayser+1986, Witt1990} as 
\begin{equation}
    \bm{w} = \bm{z} - \sum^{N}_{i=1} \frac{\epsilon_i}{\overline{\bm{z}} - \overline{\bm{a}}_i}, \label{eq:lens_eq}
\end{equation}
where $\bm{w}$ and $\bm{z}$ are the source and image positions in complex coordinates, respectively, and $\bm{a}_i$ denotes the position of the $i$-th lens. $\overline{\bm{z}}$ is defined as a conjugate of $\bm{z}$. The parameter $\epsilon_i$ represents the fractional mass of the $i$-th lens normalized by the total lens mass $M$, so that $\sum_i \epsilon_i = 1$. All coordinates are expressed in units of the Einstein radius corresponding to $M$. Because of the complex conjugation, this lens equation defines a nonlinear and non-analytic mapping from the image plane to the source plane. As a result, obtaining the image positions $\bm{z}$ for a given source position $\bm{w}$ generally requires numerical root-finding methods.

When assuming a point source, the total magnification is computed as the sum of the signed magnifications of all images. Each image contributes an amount equal to the inverse of the absolute value of the Jacobian determinant of the lens mapping $\bm{z} \mapsto \bm{w}$, evaluated at the corresponding image position $\bm{z}_j$:
\begin{equation}
    A(\bm{w}) = \sum_{j=1}^{N_{\rm image}} \frac{1}{\left| \det J(\bm{z}_j) \right|}, \label{eq:amp_point}
\end{equation}
where $N_{\rm image}$ is the number of images. The Jacobian determinant at $\bm{z}_j$ is given\footnote{The lens equation involves complex conjugation ($\overline{\bm{z}}$), making the mapping $\bm{z} \mapsto \bm{w}$ non-analytic (non-holomorphic).} by
\begin{equation}
    \det J(\bm{z}_j) = 1 - \left| \sum_{i=1}^{N} \frac{\epsilon_i}{(\overline{\bm{z}}_j - \overline{\bm{a}}_i)^2} \right|^2. \label{eq:detJ}
\end{equation}
To evaluate Equation~\eqref{eq:amp_point}, all image positions $\{\bm{z}_j\}$ corresponding to a given source position $\bm{w}$ must first be found by solving the lens equation. For multiple point-mass lenses, Equation~\eqref{eq:lens_eq} can be reformulated as a complex polynomial in $\bm{z}$:
\begin{equation}
    \mathcal{P}(\bm{z}|\bm{w}, \{\bm{a}_i\}, \{\epsilon_i\}) = \sum^{N_\mathrm{deg}}_{l=0} b_l \bm{z}^{l}=0, \label{eq:lens_eq_z}
\end{equation} 
where the polynomial degree is $N_\mathrm{deg} = 5(N - 1)$ for an $N$-lens system \citep[e.g.,][]{Rhie2003}, yielding $N_\mathrm{deg} = 5$ and $10$ for binary and triple lenses, respectively. The number of physical images ranges from $N+1$ to $5(N-1)$, but not all roots of Equation~\eqref{eq:lens_eq_z} correspond to real images. Some may be spurious or correspond to non-physical solutions, and thus each root of images must be validated by substituting it back into Equation~\eqref{eq:lens_eq}. The magnification diverges where $\det{J} = 0$, which defines the critical curves in the image plane and maps to caustics in the source plane. These regions produce strong, distinctive magnification features and are particularly sensitive to modeling errors.

\subsection{Image-Centered Ray Shooting (ICRS)}
For extended sources, the total magnification is given by the surface-brightness-weighted integral of the point-source magnification over the source disk. In the presence of caustic structures, particularly during near-caustic or caustic crossings, the magnification pattern can exhibit sharp spatial gradients, rendering the point-source approximation and analytic methods (e.g., hexadecapole expansions; \citealp{Gould2008, Cassan2017}) inaccurate. In such cases, numerical integration is required to evaluate the finite-source magnification.

In the ICRS method, the lens equation is first solved to determine the point-source image positions $\{\bm{z}_j\}$ for a given center of source position $\bm{w}_{c}$ (and arbitrary positions within the source). Around each image $\bm{z}_j$, a localized grid is constructed on the image plane, which must fully enclose the extent of the corresponding image to ensure that all contributing rays are captured. Each grid point $\bm{z}_j^{(k)}$ is mapped back to the source plane via the lens equation: $\bm{w}_j^{(k)} = f_{\rm map}(\bm{z}_j^{(k)})$, where $f_{\rm map}$ denotes the mapping defined by Equation~\eqref{eq:lens_eq}. We then define an indicator function that determines whether each mapped point lies within the angular radius $\rho$ of the source (in units of $\theta_{\rm E}$):
\begin{eqnarray}
    \chi_j^{(k)} = 
    \begin{cases}
        1 & \text{if } |\bm{w}_j^{(k)} - \bm{w}_c| < \rho, \\
        0 & \text{otherwise}.
    \end{cases}
\end{eqnarray}
The total finite-source magnification is then approximated as
\begin{eqnarray}
    A(\bm{w}_c, \rho) \approx \frac{1}{F_0} \sum^{N_{\rm image}}_j \sum^{N_{\rm grid}}_k \chi_j^{(k)} I(\bm{w}_j^{(k)}) \Delta(\bm{z}_j^{(k)}),
\end{eqnarray}
where $N_{\rm grid}$ is the number of grid points per image, $I(\bm{w})$ is the source surface brightness profile (e.g., uniform or limb-darkened), $\Delta(\bm{z}_j^{(k)})$ is the area element associated with each grid point in the image plane, and the normalization $F_0$ is the total unlensed source flux:
\[
F_0 = \int_{|\bm{w}' - \bm{w}_c| \leq \rho} I(\bm{w}') \, d^2\bm{w}'.
\]

This method enables accurate integration of the lensed surface brightness while maintaining computational efficiency by restricting the calculation to regions around the true image positions. Furthermore, the mapping $f_{\rm map}(\bm{z})$ is inherently local and independent across grid points, making the algorithm naturally amenable to parallel execution on GPUs. However, achieving high-accuracy magnification estimates requires careful treatment of image boundaries and integration schemes \citep{Bennett+2010}. We describe our implementation and associated optimizations in the following section.

\section{Implementation \label{sec:implementation}}

To enable fast and high-precision microlensing modeling on GPUs, \texttt{microJAX} implements a parallelizable version of the ICRS algorithm. All computational components are designed to be fully compatible with automatic differentiation, thereby supporting gradient-based optimization and inference techniques. In this section, we describe the key algorithmic components of \texttt{microJAX}, beginning with the numerical solution of the lens equation.

\subsection{Solving the Lens Equation}
Numerical solutions to Equation~\eqref{eq:lens_eq_z} can be obtained by casting the polynomial into an $N_{\rm deg} \times N_{\rm deg}$ companion matrix and computing its eigenvalues. Although this eigenvalue formulation guarantees all $N_{\rm deg}=5(N-1)$ roots without initial guesses, the computational cost scales as $\mathcal{O}(N_{\rm deg}^3)$, and it can suffer from numerical instability in some cases. Consequently, contemporary microlensing codes prefer iterative root-refinement schemes such as Laguerre's method and the updated version of the algorithm proposed by \citet{Skowron+2012}, which converge quadratically once a reasonable starting estimate is available.

In \texttt{microJAX}, we adopt the Ehrlich-Aberth (EA) method \citep{Ehrlich1967, Aberth1973, Fatheddin+2022}, which simultaneously refines all root estimates. The update rule is given by
\begin{equation}
    z_k^{(n+1)} = z_k^{(n)} - \left( \frac{\mathcal{P}^{\prime}(z_k^{(n)})}{\mathcal{P}(z_k^{(n)})} - \sum_{j \neq k} \frac{1}{z_k^{(n)} - z_j^{(n)}} \right)^{-1},
\end{equation}
where $\mathcal{P}(z)$ is the polynomial defined in Equation~\eqref{eq:lens_eq_z}, and $\mathcal{P}^{\prime}(z)$ is its derivative. This method extends Newton's method by introducing a repulsive term between root estimates, enhancing convergence stability and mitigating root coalescence. A key advantage of the EA method is its inherent parallelism: all root estimates are updated simultaneously at each iteration without sequential dependencies. This property makes it particularly well-suited for massively parallel execution on GPUs \citep{Weisenbach2025}. Moreover, its deterministic control flow and fixed iteration count enable efficient compilation within JAX's static computation graph, supporting just-in-time (JIT) compilation and seamless integration with automatic differentiation frameworks. 

The initial root estimates $\{z^{(0)}\}$ are placed uniformly on a circle in the complex plane, with the radius determined by Cauchy's bound based on the polynomial coefficients. This choice ensures that all true roots are enclosed and that the iteration remains numerically stable from the outset. In practice, this initialization works reliably across a wide range of lens configurations and is used as the default in \texttt{microJAX}. Importantly, the EA method’s ability to update all root estimates simultaneously at each step makes it well-suited for vectorized evaluation using \texttt{vmap} in \texttt{JAX}. This enables efficient batch evaluation of many source positions in parallel, allowing the root-finding procedure to scale effectively on GPU architectures. For instance, using an NVIDIA A100 GPU, \texttt{microJAX} can solve $\sim10^6$ 10th-degree complex polynomials, corresponding to triple-lens systems, in under one second. This level of throughput is crucial for modeling large microlensing datasets in modern inference pipelines.

\subsection{Defining Image-centered Grid Regions}

\begin{figure}
    \centering
    \includegraphics[scale=0.55]{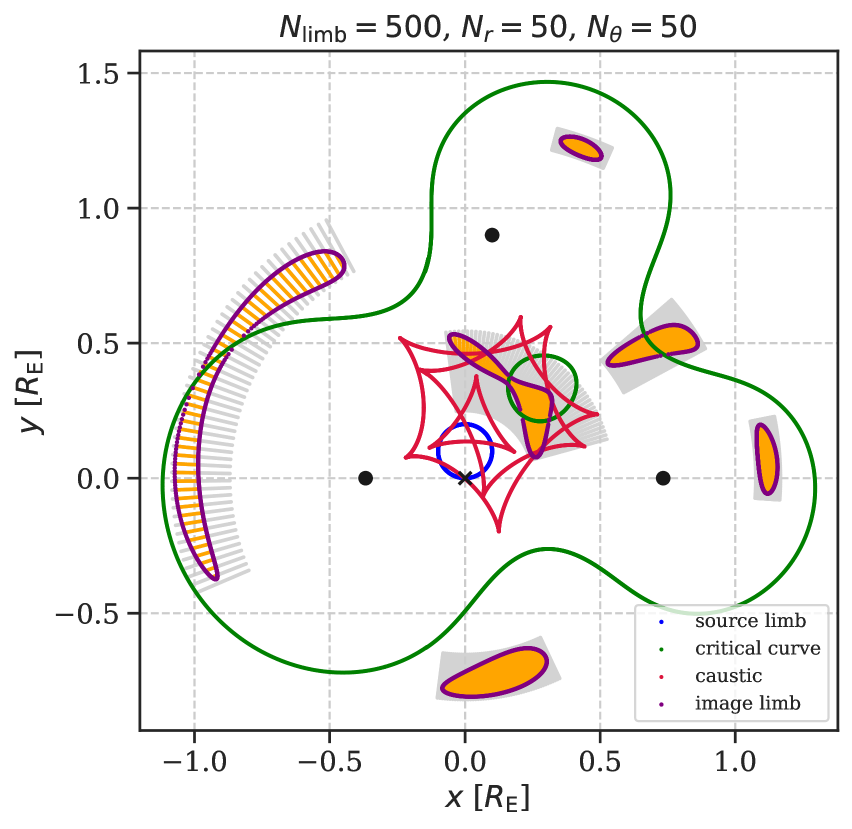}
    \caption{
    An example of a triple-lens configuration with parameters $(q, s, q_3, r_3, \psi, \rho) = (0.5, 1.1, 0.91, 1.46, 0.1)$ and source center $\bm{w}_{\rm center} = (0, 0.1)$. The red and green curves represent the caustics and critical curves, respectively, and the black dots indicate the lens positions. The source limb is sampled with $N_{\rm limb} = 500$ points (blue), which are mapped to the image plane, producing six distinct images (purple). For each image, a polar-coordinate region is defined based on the image-limb points, within which a uniform $50 \times 50$ grid is placed for integration. Grid points falling inside the source are marked in orange, while those outside are shown in gray.
    }
    \label{fig:example_grid_allocating}
\end{figure}

A key step in the ICRS algorithm is defining compact image-centered grid regions that fully enclose each lensed image while minimizing excess area to reduce computational redundancy. In traditional CPU-based implementations, such regions are often constructed by sequentially exploring the image plane. Starting from the image positions obtained by solving the lens equation for the source center $\bm{w}_c$ (and arbitrary positions within the source), inverse-ray shooting is applied iteratively to nearby points until the full image boundary is traced. While adaptive and efficient, this approach involves branching logic and sequential updates, making it poorly suited for GPU architectures and just-in-time (JIT) compilation in \texttt{JAX}.

Unlike traditional region-growing or adaptive boundary tracing methods, \texttt{microJAX} adopts a fully static and parallelizable approach to region construction. For each source position, we uniformly sample the boundary of the source with $N_{\rm limb}$ points.\footnote{This uniform sampling serves only to define integration regions for the ICRS. Our tests in various situations show $N_{\rm limb}\sim500$ is sufficient for stability. Adaptive sampling could in principle improve robustness, but its sequential nature makes it unsuitable for GPU execution.} Each of these limb points is mapped to the image plane using the lens equation, yielding a set of image-limb points that trace the extent of each image. The sampling positions on the source boundary are given by
\begin{align}
    \bm{w}^{(i)} = \bm{w}_c + \rho e^{2\pi i/N_{\rm limb}}, \quad i=0, 1, \dots, N_{\rm limb} -1
\end{align}
where $\bm{w}_c$ is the source center and $\rho$ is the angular source radius in units of the angular Einstein radius $\theta_{\rm E}$. These mapped points on the image plane define the approximate shapes and locations of the lensed images. 

Figure \ref{fig:example_grid_allocating} illustrates a result of the grid allocating algorithm for clear visualization. Each grid sector corresponds to a distinct image and is processed independently to compute the finite-source magnification. To construct the grid-locating regions efficiently, we convert the mapped image-limb points into polar coordinates $(r, \theta)$ and apply one-dimensional binning (i.e., coarse clustering) separately to the radial and angular components. This procedure identifies intervals where the image-limb points are densely populated. All combinations of radial and angular bins define candidate polar sectors in the image plane, which are tested using binary masks to determine whether they contain any image-limb points. Empty sectors are discarded, while periodic or overlapping ones (e.g., those crossing the $\theta = 0$ boundary) are merged. The total number of sectors is capped based on the maximum expected number of images. For each remaining sector, we refine the boundaries by computing the minimum and maximum values of $r$ and $\theta$ among the enclosed points and applying a small safety margin to ensure complete image coverage. The resulting compact annular sectors tightly enclose the images and are used to place integration grids. This binning-based approach scales as $\mathcal{O}(N_{\rm limb})$, in contrast to traditional adaptive methods such as boundary tracing or pairwise clustering, which typically require $\mathcal{O}(N_{\rm limb}^2)$ operations and introduce control-flow branching that hinders JIT compilation. By relying entirely on vectorized operations, our method ensures fast, scalable, and fully differentiable region construction compatible with parallel GPU execution within the JAX framework.

\subsection{Inverse-Ray Shooting and Integration Rules \label{sec:ICRS_rule}}
Given the polar grid regions defined for each image, we compute the magnification by performing inverse-ray integration within each region. For every annular sector specified by $(r_{\rm min}, r_{\rm max}, \theta_{\rm min}, \theta_{\rm max})$, we construct a fixed-resolution polar grid\footnote{Adaptive $(r,\theta)$ grids were tested but found inefficient for GPU parallelization. With our integration scheme, the uniform grid provided a better balance of speed and accuracy.} consisting of $N_r$ radial and $N_{\theta}$ azimuthal points. For each radial slice at fixed radius $r_i$, the magnification is computed by summing contributions from azimuthal segments, where the inverse-ray mapping determines whether each grid point maps back inside the source disk. At each azimuthal angle $\theta_\ell$, the corresponding grid point in the image plane is mapped back to the source plane via the lens equation, yielding the position $\bm{w}_{i,\ell}$. The Euclidean distance from the source center is computed as $d_{i,\ell} = |\bm{w}_{i,\ell} - \bm{w}_{c}|$ and is used to evaluate the surface brightness profile of the source, $I(d_{i,\ell})$. Near the image boundary, where the transition between inside and outside the source occurs, we estimate the precise edge location ($d = \rho$) by performing cubic interpolation using the distances of nearby grid points. This interpolation enables smooth and accurate integration across partially covered pixels and is particularly important when modeling limb-darkened sources. The detailed quadrature rules and interpolation formulae used in these calculations are presented in Appendix~\ref{appendix:integration}.

\subsection{Optimizing Parallel Computing and Memory Balance}

High GPU performance requires both massive parallelism and efficient memory usage. In the ICRS method, which handles extended-source effects with fine image structures, balancing computational throughput with memory constraints is critical. To address this, \texttt{microJAX} leverages JAX’s functional primitives, particularly \texttt{vmap} and \texttt{lax.scan}, to express parallelism in a compiler-friendly and memory-efficient manner. The implementation distinguishes between operations that scale well with data-level parallelism and those limited by memory or control-flow dependencies. Key parallelized components include:
\begin{itemize}[itemsep=0.0ex, topsep=0.0ex, leftmargin=*]
    \item[(i)] \textbf{Lens equation solving}: At each source-limb point, root-finding is fully vectorized with \texttt{vmap}, exploiting the embarrassingly parallel nature of the task without incurring significant memory overhead.
    \item[(ii)] \textbf{Inverse-ray shooting}: Image-plane grid points are mapped independently to the source plane. Within each sector (corresponding to a distinct image), azimuthal rays are evaluated in parallel via \texttt{vmap}, while integration across sectors is done sequentially with \texttt{lax.scan}, allowing intermediate results to be discarded and reducing memory usage. The dominant memory cost arises here, scaling with the product of radial and azimuthal grid sizes ($N_r \times N_{\theta}$).
    \item[(iii)] \textbf{Light curve modeling}: For extended-source treatment, only a preselected set of $N_{\rm ext}$ epochs (e.g., near caustics) are modeled. These are divided into batches of size $N_{\rm batch}$, processed in parallel via \texttt{vmap}, while outer iterations are handled by \texttt{lax.scan} or explicit loops. 
\end{itemize}
As a result, the memory footprint scales as $\mathcal{O}(N_r \times N_{\theta} \times N_{\rm batch})$. This memory-aware design allows modeling of thousands of time steps within realistic GPU limits. However, the use of \texttt{lax.scan}, which discards the intermediate state, precludes efficient reverse-mode automatic differentiation. Accordingly, \texttt{microJAX} currently supports only forward-mode differentiation for these components.

\subsection{Automatic Differentiation}
\begin{figure*}
    \centering
    \includegraphics[scale=0.68]{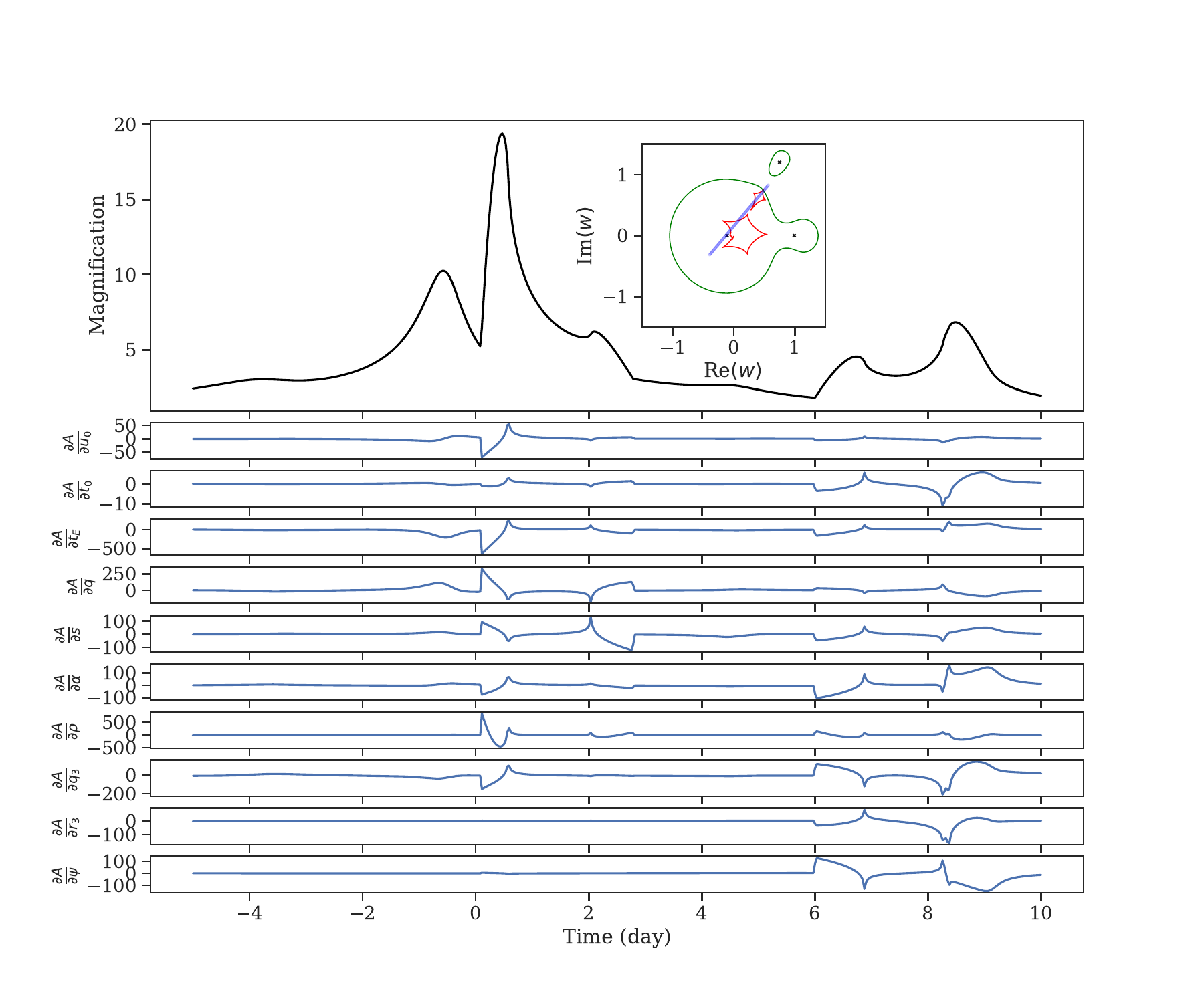}
    \caption{
    The top panel shows the triple-lens magnification curve for a uniform brightness source, with the inset displaying the corresponding caustic structure (red solid line) and source trajectory (blue solid line), where the microlensing parameters are $(t_0, t_{\rm E}, u_0, q, s, \alpha, q_3, r_3, \psi)=(0, 10\;{\rm day}, 0, 0.5, 1.1, 60^\circ, 0.03, 1.24, 76.0^\circ)$. The subsequent panels show the gradients of the magnification for each microlensing parameter. The resolutions of the inverse-ray grid are $(N_r, N_\theta)=(500, 500)$, respectively.
    }
    \label{fig:auto_diff}
\end{figure*}

\begin{figure}
    \centering
    \includegraphics[scale=0.5]{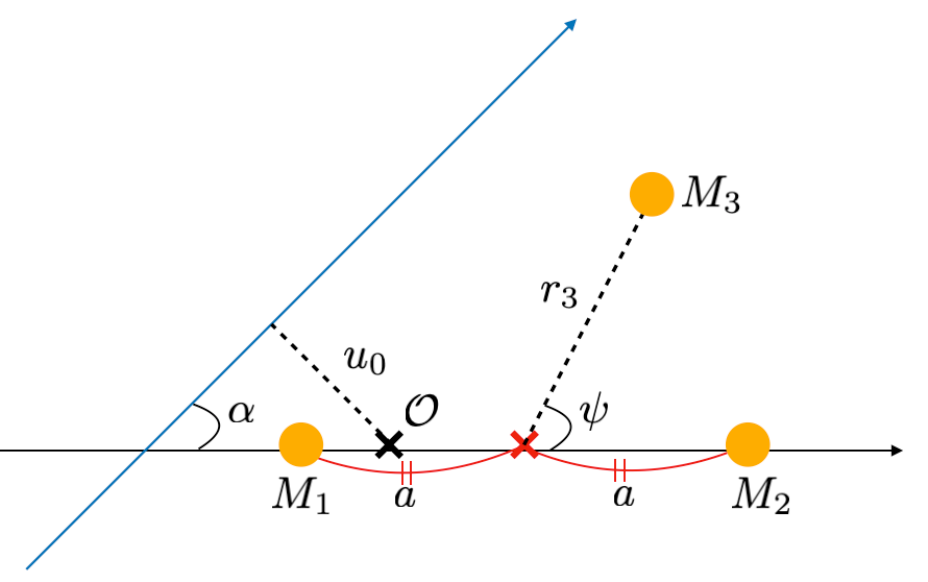}
    \caption{
    Geometrical configuration of the triple-lens system used in magnification and gradient calculations. The primary and secondary lenses ($M_1$, $M_2$) are separated by $s \equiv 2a$ along the horizontal axis, with the origin $\mathcal{O}$ defined as their barycenter. The source (blue line) moves with impact parameter $u_0$ and angle $\alpha$ relative to the binary axis. The third lens $M_3$ is placed at polar coordinates $(r_3, \psi)$ relative to the midpoint between $M_1$ and $M_2$. 
    }
    \label{fig:triple-geo}
\end{figure}

\texttt{microJAX} leverages JAX's automatic differentiation (AD) framework to compute gradients of the magnification function for model parameters. However, certain model components involve non-differentiable operations, such as Heaviside functions and conditional branches, which obstruct standard differentiation. 
We therefore define differentiable extensions of such functions with hand-written derivative rules via the \texttt{@custom\_jvp} decorator in \texttt{JAX}\footnote{JVP (Jacobian-Vector Product) is the forward-mode directional derivative; the reverse-mode counterpart is VJP (Vector-Jacobian Product), which \texttt{JAX} derives automatically from the same rule.}.

\texttt{microJAX} supports differentiable root-finding for complex polynomials via a JAX-compatible implementation of the Ehrlich-Aberth (EA) method. Consider the complex polynomial
\[
\mathcal{P}(z;\bm{b}) = \sum_{k=0}^{n} b_k\,z^{\,k},
\qquad
\bm{b}=(b_0,\dots,b_n)\in\mathbb{C}^{\,n+1},
\]
with complex variable \(z\).
Its roots
\(\bm{z}^{\ast}(\bm{b})=(z^{\ast}_{1},\dots,z^{\ast}_{n})\)
are implicitly defined by
\(\mathcal{P}(z^{\ast}_{i};\bm{b})=0\).
To quantify how each root changes with the coefficients, differentiate the identity \(\mathcal{P}(z^{\ast}_{i}(\bm{b});\bm{b})=0\) with respect to \(\bm{b}\):
\[
\frac{\partial\mathcal{P}}{\partial z}\bigl(z^{\ast}_{i};\bm{b}\bigr)
\;\frac{\mathrm{d}z^{\ast}_{i}}{\mathrm{d}\bm{b}}
\;+\;
\frac{\partial\mathcal{P}}{\partial\bm{b}}\bigl(z^{\ast}_{i};\bm{b}\bigr)
\;=\;0.
\]
Hence
\[
\boxed{\;
\frac{\mathrm{d}z^{\ast}_{i}}{\mathrm{d}\bm{b}}
\;=\;
-\,\frac{\partial\mathcal{P}/\partial\bm{b}\bigl(z^{\ast}_{i};\bm{b}\bigr)}
        {\partial\mathcal{P}/\partial z\bigl(z^{\ast}_{i};\bm{b}\bigr)}
\;}
\quad\in\mathbb{C}^{1\times(n+1)}.
\]
Here  
\(\partial\mathcal{P}/\partial\bm{b}=[(z^{\ast}_{i})^{n},(z^{\ast}_{i})^{n-1},\dots,1]\)  
is the monomial vector evaluated at \(z^{\ast}_{i}\), and  
\(\partial\mathcal{P}/\partial z = \sum_{k=1}^{n}k\,b_k\,z^{\,k-1}\) is the ordinary partial derivative.  
Stacking the gradients \(\mathrm{d}z_i^*/\mathrm{d}\bm{b}\) for all roots yields the Jacobian  
\(\mathrm{d}\bm{z}^*/\mathrm{d}\bm{b} \in \mathbb{C}^{n \times (n+1)}\),  
which characterizes the full sensitivity of the root vector to the coefficients. This closed-form expression can be registered as a custom JVP rule, so gradients propagate through the roots without back-propagating through the iterative EA updates, yielding stable and efficient gradient computation in \texttt{JAX}.

Binary functions used for inclusion tests, such as determining whether a ray falls within the source radius, are implemented using the Heaviside function $\Theta(x)$. To enable smooth automatic differentiation, these functions are overridden by sigmoid-based approximations during gradient computation. In particular, the indicator function $\chi = \Theta( \rho-d)$ used for segment classification is replaced in its JVP rule by a smooth sigmoid $\sigma(x)$ with 
\begin{eqnarray}
\frac{\partial \chi}{\partial x} = \sigma(x)(1 - \sigma(x)) \times S_{\mathrm{fac}},
\end{eqnarray}
where $S_\text{fac}$ is an empirically chosen steepness parameter controlling gradient sharpness and numerical stability. The correction factor $f(\delta)$ detailed in Appendix~\ref{appendix:integration}, used for integrating across azimuthal segments that straddle the source boundary in limb-darkened models, is defined piecewise depending on $\delta_c$. While the function itself retains its original piecewise definition during primal evaluation, its gradient is manually defined as a fixed smooth expression, typically the derivative of the second-order formula, to ensure continuity in the AD flow near the threshold. This avoids discontinuities in the AD flow that would otherwise result from switching integration rules during evaluation.

Figure~\ref{fig:auto_diff} presents the magnification curve for a triple-lens system with a uniform-brightness source, along with the gradients of magnification for each model parameter. The computed gradients reveal regions of high sensitivity, particularly near caustic crossings, demonstrating \texttt{microJAX}'s ability to evaluate both magnification and its parameter derivatives in complex microlensing models. Here the microlensing model is parameterized by the time of closest approach to the barycenter of the first and second lenses $t_0$, the impact parameter $u_0$, the Einstein radius crossing time $t_{\rm E}$, the binary mass ratio $q$, the projected separation in units of the Einstein radius $s$, the angle between the source trajectory and the binary axis $\alpha$, and the normalized source radius $\rho$. For the third lens, three additional parameters are introduced: $q_3$, the mass ratio to the primary lens; $r_3$, the projected separation between the binary midpoint and the third lens; and $\psi$, the angle between the binary axis and the direction to the third lens. The geometric definitions are illustrated in Figure~\ref{fig:triple-geo}.

\section{Performance Benchmarks \label{sec:benchmark}}
\begin{figure*}
    \centering
    \includegraphics[scale=0.55]{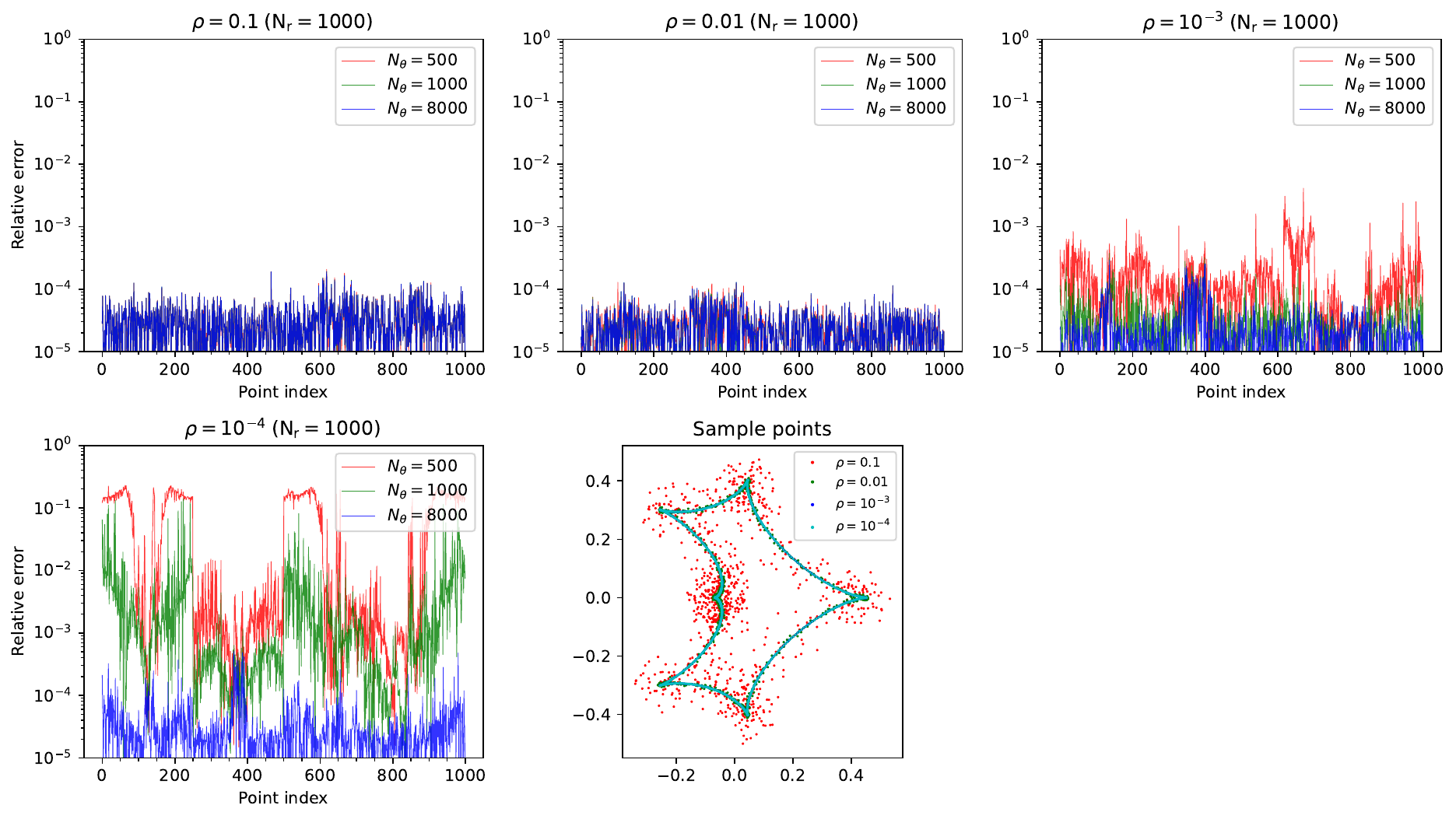}
    \caption{
    Relative accuracy of \texttt{microJAX} magnification estimates compared to \texttt{VBBinaryLensing}, as a function of source size ($\rho=0.1$ to $10^{-4}$) and angular resolution ($N_\theta = 500, 1000, 8000$) with fixed $N_r=1000$. Each panel shows the relative error in magnification computed by \texttt{microJAX}, to the reference values from \texttt{VBBinaryLensing}, for 1000 randomly sampled source positions near the central caustic of a binary lens with $q=0.1$ and $s=1$. A uniform-brightness source is assumed, and \texttt{VBBinaryLensing} is run at a relative accuracy of \(10^{-5}\). The lower-right panel illustrates the spatial distribution of test points near the caustic.
    }
    \label{fig:accuracy_compare}
\end{figure*}

\begin{figure}
    \centering
    \includegraphics[scale=0.35]{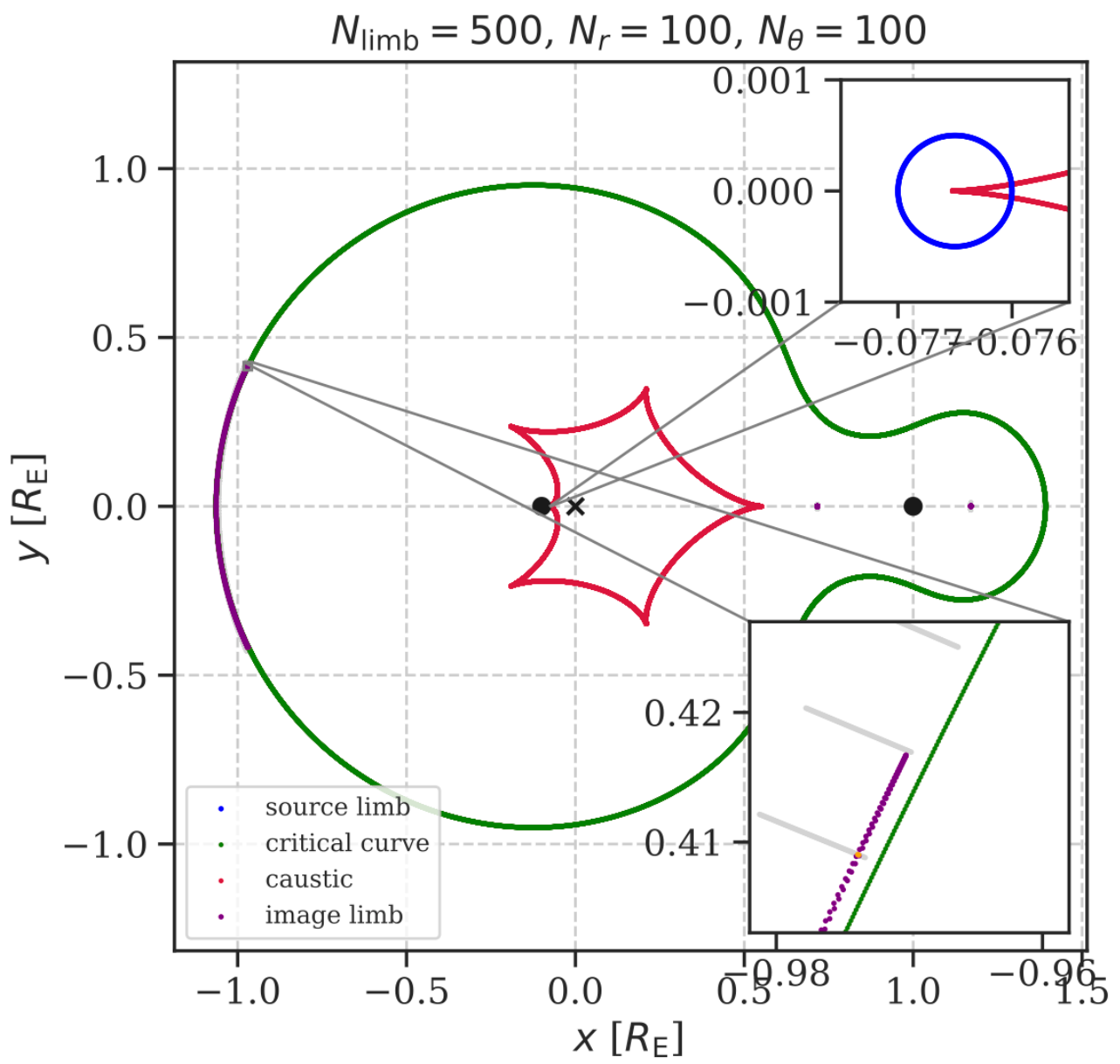}
    \caption{
    An example of a lensing geometry, similar with Figure~\ref{fig:example_grid_allocating}, where the error in magnification increases, with parameters $(q, s, \rho) = (0.1, 1.1, 5\times10^{-4})$ and $\bm{w}_{\rm center}=(-0.0765, 0)$. Each inset panel shows a zoomed-in view of each position. In this case, the angular resolution of the grid is too sparse to accurately determine the image boundary compared to the source size, illustrated in the lower-right panel.
    }
    \label{fig:erronous_example}
\end{figure}

\subsection{Accuracy Validation}
To assess the accuracy of \texttt{microJAX}, we benchmark its magnification calculations against the widely used \texttt{VBBinaryLensing} code \citep{Bozza2010, Bozza+2018, Bozza+2021}. All comparisons are made using version~3.7.0 of \texttt{VBBinaryLensing} to ensure consistency and reproducibility. Figure~\ref{fig:accuracy_compare} shows the relative error of \texttt{microJAX} with respect to \texttt{VBBinaryLensing} for source radii $\rho = (0.1, 0.01, 10^{-3}, 10^{-4})$ and azimuthal resolutions $N_{\theta}=(500,\,1000,\,8000)$. The radial resolution is fixed at \(N_{r}=1000\), a value that limits radial-integration errors to $\lesssim10^{-4}$. A uniform-brightness source is assumed\footnote{We also tested limb-darkened sources and confirmed that the conclusions remain consistent.}, and the reference code is run at a nominal accuracy of $10^{-5}$. To construct the test dataset, source centers are randomly distributed around the central caustic of a binary lens with parameters $q = 0.1$ and $s = 1$. Specifically, each test point is sampled within a circular region of radius \(\rho\) centered at a randomly selected point along the caustic curve. This ensures that all sources lie within a distance \(\rho\) from the caustic, effectively probing regions of strong magnification variation. The sampling configuration is illustrated in the lower-right panel of Figure~\ref{fig:accuracy_compare}. We fix the number of limb points to \(N_{\mathrm{limb}} = 500\), which is sufficiently high to ensure that limb discretization does not significantly affect the relative magnification error.

For large sources (\(\rho \gtrsim 10^{-2}\)), all configurations achieve relative errors below \(10^{-4}\). As the source size decreases near the caustic, the image boundary significantly elongates primarily in the angular direction while remaining relatively comparable in the radial direction. Consequently, accurately tracing that elongated boundary requires a finer azimuthal grid $(N_\theta)$, whereas the radial grid rarely misses the boundary. Figure \ref{fig:erronous_example} illustrates a case where a coarse angular grid determines an inaccurate image boundary compared to the source size, producing sizeable errors in magnification calculation. When the integrand in the radial direction is smooth, the quadrature error scales as $N_{r}^{-2}$ and is averaged over the entire source area; thus, using coarse $N_{r}$ only gradually degrades the precision. On the other hand, azimuthal sampling is far more sensitive: any boundary determination directly impacts the magnification, so insufficient $N_\theta$ alone can introduce spikes and discontinuities in the light curve even if $N_r$ is adequate. In the most demanding case (\(\rho = 10^{-4}\)), only the highest resolution tested (\(N_{\theta}=8000\)) keeps the relative error below \(10^{-3}\), underscoring the need for dense angular sampling in small-source regimes.\footnote{The total angular extent of a lensed image is at most of order $\pi$. Therefore, to achieve an angular resolution comparable to the source radius $\rho$, the required number of azimuthal grid points is of order $\sim 1/(\pi \rho)$.} Although increasing $N_{\theta}$ raises computational cost and memory per epoch, the total cost is offset in practice by the shorter source-crossing time $t_\star = \rho\, t_{\rm E}$, which limits the number of epochs where extended-source effects must be evaluated. As a result, the effective computational burden remains manageable even at high angular resolution.

\subsection{Computational Performance}
\begin{figure}
    \centering
    \includegraphics[scale=0.45]{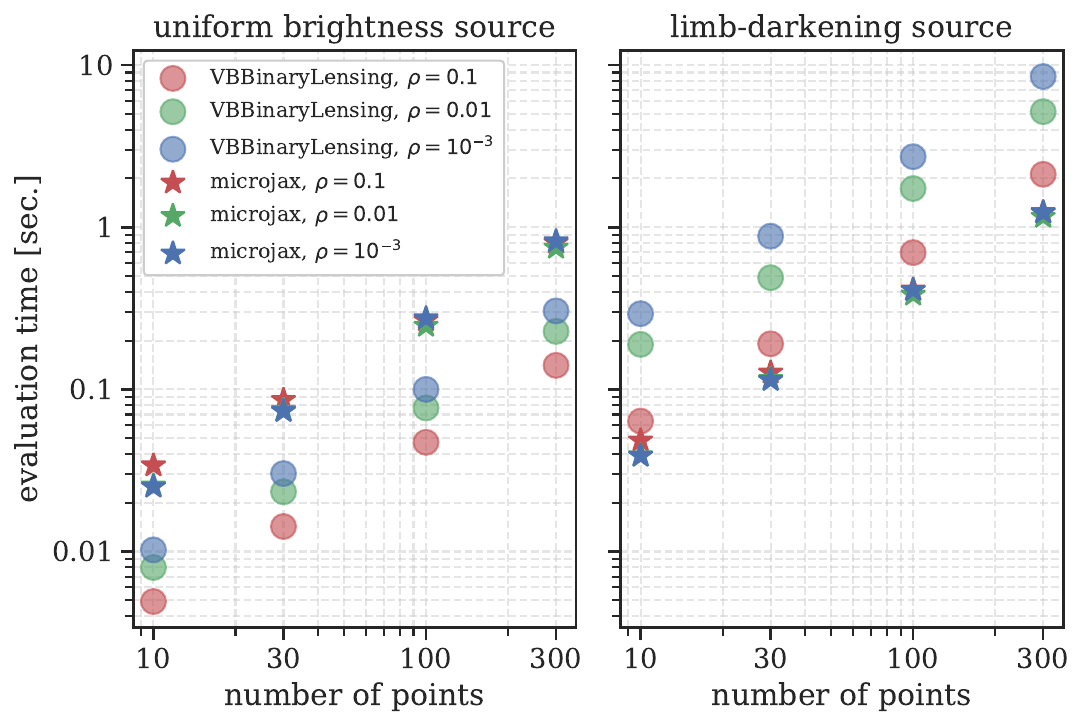}
    \caption{
    Total computation time per magnification evaluation is shown as a function of the number of light curve points (epochs) for both uniform (left) and limb-darkened (right) source profiles. The results compare \texttt{microJAX} (stars) and \texttt{VBBinaryLensing} (circles) across different source sizes $\rho$. We set the relative error in \texttt{VBBinaryLensing} to $10^{-4}$, and adopt $(N_r, N_\theta)=(10^{3}, 10^{3})$ in \texttt{microJAX}. All benchmarks were performed using an AMD EPYC 7452 CPU (2.35 GHz base frequency, single-threaded) for \texttt{VBBinaryLensing}, and an NVIDIA A100 PCIe GPU (40 GB) for \texttt{microJAX}. All timings were measured with JIT compilation enabled and exclude compilation overhead.
    }
    \label{fig:timing_comparison}
\end{figure}

Figure~\ref{fig:timing_comparison} compares the total computation time per magnification evaluation as a function of the number of light curve points (epochs), for both \texttt{VBBinaryLensing} and \texttt{microJAX}, under uniform and limb-darkened source assumptions. Both methods exhibit good scalability with the number of epochs, but the computation time for each point reflects the fundamental differences in how magnification is evaluated. \texttt{VBBinaryLensing} employs a contour integration method that is highly optimized for uniform sources, achieving performance up to $\sim10\times$ faster than in the limb-darkened case, where brightness weighting and adaptive sampling introduce significant computational overhead. In contrast, \texttt{microJAX} shows minimal difference (less than a factor of 2) between uniform and limb-darkened cases.  This efficiency arises from its use of a static computation graph, which allows all operations to be compiled efficiently ahead of time. Consequently, the total runtime is nearly independent of the source size~$\rho$. While \texttt{microJAX} can be up to $\sim$4-5$\times$ slower than \texttt{VBBinaryLensing} for large, uniform sources, it significantly outperforms in the small-source, limb-darkened regime, achieving 5-6$\times$ speedup due to its stable performance profile.

It should be noted, however, that a direct comparison between \texttt{VBBinaryLensing} (running on CPU) and \texttt{microJAX} (executed on GPU) is inherently limited by differences in hardware architecture and parallelization strategies. The reported speed differences should therefore be interpreted as relative performance under typical usage scenarios, rather than absolute benchmarks. 

Overall, these results demonstrate that \texttt{microJAX} offers competitive performance and superior scaling for complex source models, particularly in the small-source, limb-darkened regime most relevant for high-precision microlensing analyses.

\begin{deluxetable}{lcccc}          
\tablecaption{Public Codes of Microlensing Magnification Calculation\label{tab:codes}}
\tablehead{
  \colhead{Codes} & \colhead{Method\tablenotemark{*}} & \colhead{Max $N_{\rm lens}$} & \colhead{GPU} & \colhead{Differentiable}
}
\startdata
\texttt{microJAX}        & ICRS & $3$          & \cmark & \cmark \\
\texttt{eesunhong}       & ICRS & $2$          & \xmark & \xmark \\
\texttt{VBBinaryLensing} & CI   & $2$          & \xmark & \xmark \\
\texttt{VBMicrolensing}  & CI   & $\ge3$       & \xmark & \xmark \\
\texttt{Twinkle}         & CI   & $2$          & \cmark & \xmark \\
\texttt{caustics}        & CI   & $3$          & \tmark & \cmark \\
\texttt{microlux}        & CI   & $2$          & \tmark & \cmark \\
\enddata
\tablenotetext{*}{
ICRS = Image-centred Ray Shooting method. CI = Contour Integrating method.
}
\tablecomments{
\cmark = Supported; \xmark = Not supported; \tmark = JAX backend available, but no dedicated GPU optimization.
All listed codes are designed to compute microlensing magnification as a function of source position and lens configuration. \texttt{microJAX}\footnote{\url{https://github.com/ShotaMiyazaki94/microjax}}: This work. \texttt{eesunhong}\footnote{\url{https://github.com/golmschenk/eesunhong}}: CPU-optimized ICRS \citep{Bennett+1996, Bennett+2010}. \texttt{VBBinaryLensing}\footnote{\url{https://github.com/valboz/VBBinaryLensing}}: CPU-optimized contour integrating for binary-lens \citep{Bozza2010, Bozza+2018}. \texttt{VBMicrolensing}\footnote{\url{https://github.com/valboz/VBMicrolensing}}: CPU-optimized contour integrating for multiple-lens \citep{Bozza+2025}. \texttt{Twinkle}\footnote{\url{https://github.com/AsterLight0626/Twinkle}}: GPU-optimized contour integrating for binary-lens \citep{Wang+2025}. \texttt{caustics}\footnote{\url{https://github.com/fbartolic/caustics}}: differentiable contour integrating \citep{Bartolic2023}. \texttt{microlux}\footnote{\url{https://github.com/CoastEgo/microlux}}: differentiable \texttt{VBBinaryLensing} \citep{Ren+2025}.
}
\end{deluxetable}

\subsection{Comparison with Other Microlensing Codes}

Table~\ref{tab:codes} summarizes key features of publicly available codes for computing microlensing magnification, including their numerical methods, GPU support, and differentiability. Among them, \texttt{microJAX} uniquely combines differentiability, GPU acceleration, and support for triple-lens configurations, representing capabilities not simultaneously realized in other tools. Unlike other well-established codes, such as \texttt{eesunhong} \citep{Bennett+1996, Bennett+2010} and \texttt{VBBinaryLensing} \citep{Bozza2010, Bozza+2018}, which are non-differentiable and thus unsuitable for gradient-based inference, \texttt{microJAX} provides a fully differentiable framework tailored for modern probabilistic modeling. Although differentiable alternatives such as \texttt{caustics} \citep{Bartolic2023} and \texttt{microlux} \citep{Ren+2025} exist, they are not optimized for complex source profiles. To address this gap, \texttt{microJAX} implements a GPU-accelerated, batched ICRS algorithm that maintains sub-pixel accuracy while supporting arbitrary source profiles. This is not merely a technical optimization but a critical enabler: high-dimensional Bayesian inference demands flexible, physically accurate models. \texttt{microJAX} preserves the differentiability of such models without compromising computational tractability, ensuring coherence between model expressiveness and inference efficiency.

CPU- and GPU-based frameworks may play complementary roles in microlensing inference workflows. CPU-oriented tools are well-suited for large-scale exploratory analyses, such as grid-based searches over a wide range of initial conditions using CPU server clusters, serving as effective guides toward high-precision solutions. In particular, approximate posterior modes identified using CPU-based auto-differentiable frameworks may inform subsequent refinement using \texttt{microJAX}, which is optimized for GPU-accelerated modeling in the final stages of parameter inference. This potential complementarity suggests the importance of continued development of both types of tools to enable scalable and flexible analysis pipelines tailored to different phases of model evaluation.

\section{Applications \label{sec:application}}

\begin{deluxetable}{lcc}        
\tablecaption{Injection-Recovery Test Setting and Results\label{tab:test_injection}}
\tablehead{
  \colhead{Parameters} & \colhead{Injected} &
  \colhead{\begin{tabular}{c}Recovered\\(median \& 68\% CI)\end{tabular}}
}
\startdata
$t_0$                               & $0$       & $0.001^{+0.001}_{-0.002}$\\
$\log{(t_{\rm E}/\mathrm{day})}$    & $1.21$    & $1.209^{+0.004}_{-0.004}$\\
$u_0$                               & $0.06$    & $0.060^{+0.001}_{-0.001}$\\
$\log{q}$                           & $-5.0$    & $-5.04^{+0.05}_{-0.04}$\\
$\log{s}$                           & $0.012$   & $0.012^{+0.001}_{-0.002}$\\
$\alpha$                            & $4.156$   & $4.153^{+0.002}_{-0.002}$\\
$\log{\rho}$                        & $-2.74$   & $-2.74^{+0.01}_{-0.02}$\\
\enddata
\tablecomments{%
Posterior sampling is performed using HMC with 500 warm-up steps and 5,000 retained samples, executed on an NVIDIA A100 GPU. The full run required $\sim 19$ hours, corresponding to an average of 12.5\,second per retained sample. Each proposal was generated with 15 leapfrog steps, and the mean acceptance probability was $\geq$ 90\%.
}
\end{deluxetable}

\begin{figure*}
    \centering
    \includegraphics[scale=0.9]{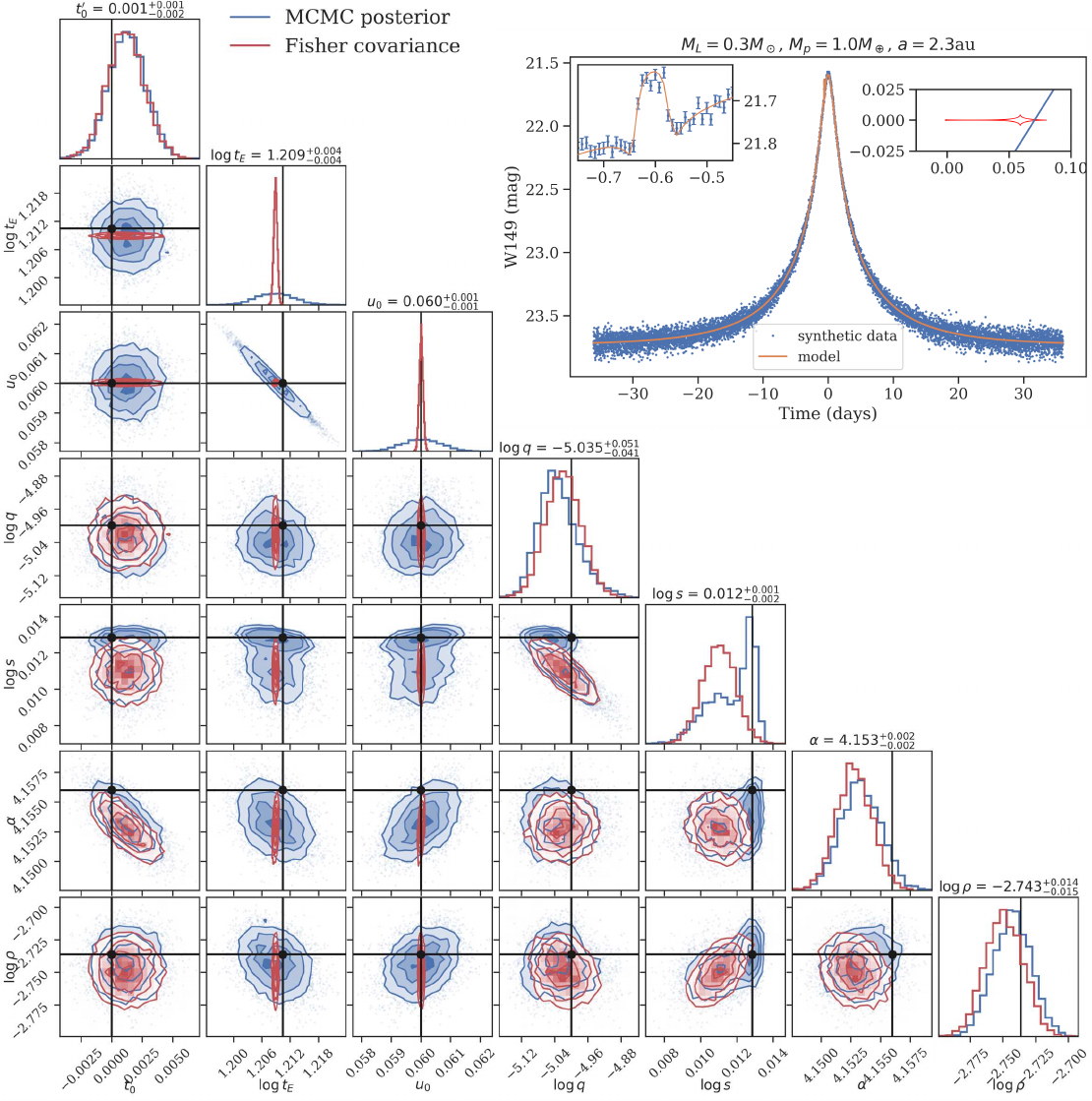}
    \caption{
    The top right panel shows the synthetic light curve observed by the Roman, where blue error bars are $W149$-band measurements and the orange solid line is the model curve. The top-right inset in the same panel displays the caustic geometry (red solid line) and the source trajectory (blue line). The corner plot shows the posterior distributions of the model parameters (blue) along with the Fisher covariance contours (red) evaluated at the best-fit parameters obtained with gradient-based optimization before inference. The ground truths (the injected parameters) are shown as black cross-hatched points.
    }
    \label{fig:roman_mcmc}
\end{figure*}

\begin{figure*}
    \centering
    \includegraphics[scale=0.8]{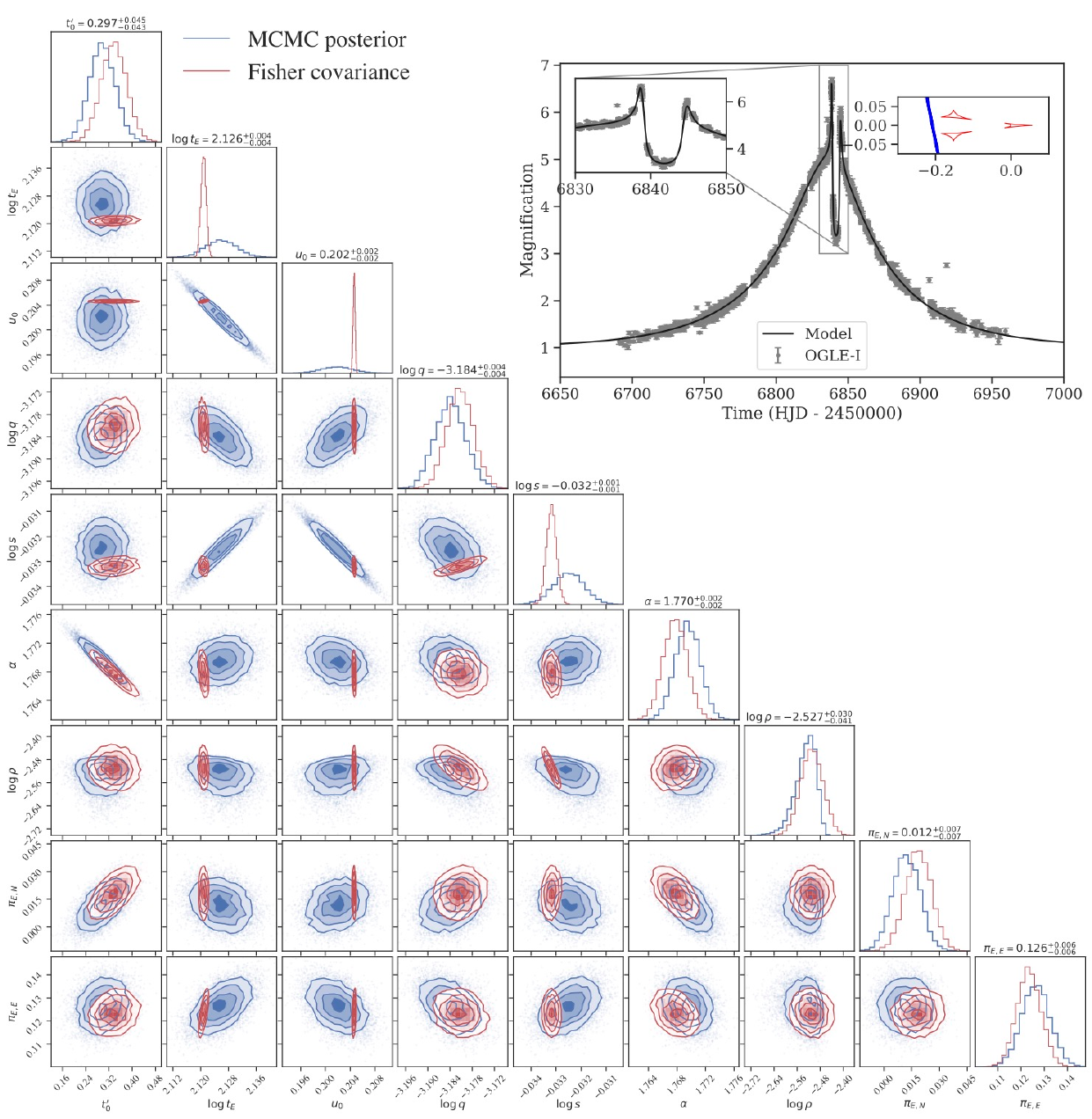}
    \caption{
    The top right panel shows the observed light curve of OGLE-2014-BLG-0124. Gray error bars indicate the $I$-band photometry from the OGLE-IV survey, and the black solid line denotes the best-fit binary-lens model including orbital parallax. The inset in the same panel displays the caustic geometry (red solid line) and the source trajectory (blue line). The corner plot shows the posterior distributions of the model parameters (blue) along with the Fisher covariance contours (red), evaluated at the best-fit solution obtained with gradient-based optimization before inference.
    }
    \label{fig:ob140124}
\end{figure*}

This section demonstrates the practical capabilities of \texttt{microJAX} through two complementary experiments. First, we conduct a parameter recovery test using a synthetic Roman-like light curve with known ground-truth parameters. Second, we apply the method to the well-studied planetary microlensing event OGLE-2014-BLG-0124, utilizing real and heterogeneous photometric data. Together, these experiments illustrate the statistical reliability and robustness of \texttt{microJAX} under both idealized and realistic observational conditions. Prior to posterior sampling, we perform gradient-based optimization using the \texttt{optax} library \citep{deepmind2020jax} to obtain a good initial estimate. Throughout the analysis, we employ posterior sampling via Hamiltonian Monte Carlo (HMC) with the No-U-Turn Sampler (NUTS; \citealt{Duane+1987, Betancourt2017}) as implemented in \texttt{NumPyro} \citep{Phan+2019, bingham2019pyro}.

\subsection{Parameter Recovery Test with a Synthetic Roman-like Event}
To assess inference accuracy under controlled conditions, we generate a 72-day synthetic light curve sampled at a 12-minute cadence, adopting fiducial microlensing parameters. Gaussian photometric uncertainties are independently assigned based on the $W149$-band magnitude-uncertainty relation presented in Figure 4 of \citet{Penny+2019}. The light curve includes a planetary perturbation induced by a lens system in which an Earth-mass planet orbits an M-dwarf host. The injected microlensing parameters are set as $\bm{\phi} \equiv (t_0, \log({t_{\rm E}/{\rm day}}), u_0, \log{q}, \log{s}, \alpha, \log{\rho}) = (0., 1.21, 0.06, -5.0, 0.012, 4.156, -2.74)$. 

Instead of placing priors directly on the physical parameters, we reparameterize the model using a latent standard normal variable. 
Specifically, we draw a latent vector $\bm{\beta} \sim \mathcal{N}(\bm{0}, \bm{I})$ and apply an element-wise affine transformation to obtain the physical parameters: $\bm{\phi}=\bm{\mu} + \bm{\sigma} \odot \bm{\beta}$, where $\bm{\mu}$ is the maximum a posterior (MAP) estimate obtained via gradient-based optimization, and $\bm{\sigma}$ is a vector of scale parameters, given by the square roots of the diagonal elements of the Fisher covariance matrix evaluated at $\bm{\mu}$ and inflated by a factor of 30. The symbol $\odot$ denotes the Hadamard (element-wise) product. This approach corresponds to an independent Gaussian prior in the physical space:
\begin{align}
    \phi_i \sim \mathcal{N}(\phi_{i, {\rm init}}, 30 \cdot\sigma_{i, {\rm Fisher}}), 
\end{align}
where $\phi_{i, {\rm init}}$ is the MAP estimate, and $\sigma_{i, {\rm Fisher}}$ is the corresponding diagonal element of the Fisher covariance matrix evaluated at that point. While the full Fisher matrix encodes local parameter correlations, we deliberately ignore off-diagonal terms to avoid injecting potentially unreliable structure into the prior. The diagonal approximation with variance inflation provides a conservative, weakly informative prior that respects the local scale of each parameter without overcommitting to possibly inaccurate correlation directions. In other words, this is equivalent to independent Gaussian priors centered on the MAP values with inflated Fisher variances. The reparameterization with latent standard normal variables is introduced only for computational convenience, ensuring stable gradient-based inference.

The microlensing light curve is modeled as
\begin{equation}
F_{\rm obs}(t) = \left[A(t; \bm{\phi}) -1\right] F_{S} + F_{\rm base},
\end{equation}
where $A(t; \bm{\phi})$ is the model magnification as a function of time and microlensing parameters, and $F_{S}$ and $F_{\rm base}$ denote the source and baseline fluxes, respectively. Given that the photometric uncertainty of the light curve reaches $\sigma_{W149} \sim 0.01$, we adopt integration resolutions of $(N_r, N_\theta) = (10^3, 10^3)$, which conservatively ensures an accuracy better than $10^{-4}$. Since the flux parameters enter linearly, they are analytically marginalized in the likelihood evaluation, as described in Appendix~\ref{appendix:marginalize}. Posterior sampling is performed using HMC with 500 warm-up steps and 5,000 retained samples, executed on an NVIDIA A100 GPU. Table \ref{tab:test_injection} shows the settings and results of the injection-recovery test.\footnote{For reproducibility, the code used to perform this test is available at \url{https://github.com/ShotaMiyazaki94/test_microjax}.}

Figure \ref{fig:roman_mcmc} presents the synthetic Roman-like light curve (top right panel) alongside a corner plot summarizing the HMC posterior. Splitting the single HMC chain yields Gelman-Rubin diagnostics \citep{Gelman+1992} of $\hat{R}<1.00$ and effective sample sizes exceeding 800 for all parameters, indicating convergence well. In the corner plot, red ellipses represent a Gaussian approximation based on the Fisher matrix evaluated at the best-fit point. These ellipses agree with the posterior for well-constrained parameters such as the $(t_0, \alpha)$ plane. Although all injected values (black cross-hairs) are recovered within the 90\% highest-density regions of the posterior, some Fisher ellipses entirely fail to enclose them. This discrepancy is particularly evident in poorly constrained or degenerate directions, such as \((\log t_{\rm E}, u_0)\) and \((\log q, \log s)\), where the posterior exhibits strong skewness or curvature. In such cases, the Fisher approximation not only underestimates the uncertainty but also misrepresents the correlation structure, with both scale and orientation deviating significantly from the true posterior geometry. By contrast, the HMC-based posterior fully captures these non-Gaussian features, accurately recovering the injection values even in the presence of pronounced degeneracies. This highlights the robustness of \texttt{microJAX} with HMC in resolving complex parameter structures beyond the reach of local Gaussian approximations.

\subsection{Real Planetary Event: OGLE-2014-BLG-0124}
We further demonstrate the applicability of \texttt{microJAX} to real data by analyzing the well-characterized planetary microlensing event OGLE-2014-BLG-0124 \citep{Udalski+2015a}. We use only the $I$-band photometry from the OGLE-IV survey \citep{Udalski+2015b}, retrieved from the NASA Exoplanet Archive \citep{Christiansen+2025}, and fit the light curve using a binary-lens model that includes the orbital parallax effect. The model consists of nine parameters, defined as $\bm{\phi} \equiv (t_0, \log({t_{\rm E}}/{\rm day}), u_0, \log{q}, \log{s}, \alpha, \log{\rho}, \pi_{{\rm E}, N}, \pi_{{\rm E}, E})$, where $\pi_{{\rm E}, N}$ and  $\pi_{{\rm E}, E}$ are the north and east components of the microlens parallax vector \citep{Gould2004}, respectively. This event has degenerate solutions of $u_0>0$ and $u_0<0$, but here we focus on the $u_0>0$ solution. During HMC, we also infer a flux error rescaling factor $f_{\rm err}$, defined such that the rescaled photometric uncertainties are $\sigma^{\prime}_{i} = f_{\rm err} \sigma_{i}$ where $\sigma^{\prime}_{i}$ and $\sigma_{i}$ are the rescaled and original uncertainties, respectively. Posterior sampling is performed using HMC with 500 warm-up steps and 10,000 retained samples in a single chain. The overall inference procedure follows the same framework as applied in the synthetic test case.

Figure~\ref{fig:ob140124} shows the observed light curve (upper right panel) and the resulting corner plot from parameter inference using HMC. The posterior of the error rescaling factor is $f_{\rm err}=1.711^{+0.013}_{-0.013}$ (68\% credible interval), and, for all the parameters, $\hat{R}<1.00$ and the ESS exceeds 5,000. Since our light curve and model\footnote{For example, our model does not account for lens orbital motion.} differ from those of \citet{Udalski+2015a}, we do not attempt a direct comparison of the inferred parameters. Nevertheless, our results are qualitatively consistent with the original study, recovering a comparable planetary signal and clear parallax detection. This supports the reliability of our framework under realistic observational conditions. As in the synthetic test, the Fisher matrix does not fully capture the posterior geometry for several parameters in the real-data analysis. These discrepancies imply the limitations of the Fisher-based prior with correlation.

\subsection{Caveats and Lessons Learned}
Our experiments confirm that HMC enables robust and efficient posterior sampling, even in high-dimensional microlensing models. However, several caveats must be noted. The Fisher information matrix, while useful for setting initial parameter scales, often fails to capture the true posterior geometry, particularly in the presence of strong degeneracies, skewness, or multimodal structure. The resulting Gaussian approximation typically underestimates uncertainties and distorts parameter correlations, leading to error ellipses that may misalign with the true posterior and fail to enclose the ground-truth parameters. When reparameterization is performed using the Fisher matrix, these inaccuracies propagate into the prior. Because HMC operates in the latent space defined by this transformation, misalignment between the Fisher-based prior and the posterior can suppress sampling efficiency and hinder exploration of relevant modes. We observe these effects even under idealized conditions, suggesting that full-covariance Fisher priors \citep[e.g.,][]{Ren+2025} are only valid when the posterior is approximately Gaussian.

These issues are especially pronounced in microlensing, where model degeneracies and parameter correlations are not rare. Even simplified assumptions, such as diagonal priors, may also risk overlooking critical posterior structure. While Fisher-based priors can serve as a practical and scalable starting point, we advocate for flexible prior models and rigorous posterior diagnostics, especially in complex or weakly constrained inference problems.

\section{Discussion and Conclusion} \label{sec:conclusion}
We have presented \texttt{microJAX}, a differentiable microlensing modeling framework that combines GPU-accelerated, image-centered inverse-ray shooting with JAX-based automatic differentiation. Designed for scalability and physical fidelity, \texttt{microJAX} supports gradient-based inference in extended-source and multiple-lens models. Benchmarks against \texttt{VBBinaryLensing} show that \texttt{microJAX} achieves sub-second times of extended-source magnification evaluation for hundreds of epochs on a single GPU while maintaining relative magnification error less than $10^{-4}$, making it suitable for precise and high-throughput applications. Case studies further demonstrate its ability to recover model parameters efficiently under both idealized and observationally realistic conditions.

Despite its current capabilities, several key extensions remain for future work. The framework currently relies on forward-mode automatic differentiation, which scales linearly with the number of parameters and becomes costly in high-dimensional models. Reverse-mode differentiation, which offers constant-cost gradient evaluation with respect to parameter dimensionality \citep{Blondel+2024}, is not yet implemented due to memory constraints: storing intermediate values from the inverse-ray shooting process exceeds typical GPU memory limits. We plan to address this through checkpointing strategies \citep{Griewank01011992}, enabling memory-efficient reverse-mode differentiation and significantly faster gradient computation.

Support for higher-order microlensing effects, such as lens orbital motion \citep{Dominik+1998, Bennett+2010LOM, Skowron+2011}, and xallarap \citep{Bennett+2008, Miyazaki+2020, Miyazaki+2021, Rota+2021}, as well as an increased number of lens masses \citep{Bozza+2025}, is also a priority. Furthermore, high-resolution imaging surveys such as Roman will enable the detection of astrometric microlensing signals, offering complementary constraints on lens properties \citep{Gould+2014, Wyrzykowski+2020, Lam+2022, Sahu+2022}. Incorporating such astrometric data into the modeling framework will be essential for achieving high-precision, fully-consistent inferences. These extensions, while critical, significantly increase model complexity and parameter dimensionality. As such, automatic differentiation will remain central to enabling scalable inference.

An additional direction for future development involves the use of more realistic noise models. The assumption of independent Gaussian errors breaks down in many practical cases, particularly in precise space-based observations such as those from \textit{Kepler} \citep[e.g.,][]{Carter+2009, Stumpe+2012}. Incorporating time-correlated and heteroscedastic uncertainties will be critical for robust parameter inference. The growing ecosystem of JAX-based probabilistic tools, including differentiable Gaussian processes and state-space models \citep[e.g.,][]{Pinder2022, tinygp+2024, gallifrey+2025}, provides a promising foundation for integrating physical modeling and noise characterization within a unified, end-to-end framework.

In summary, \texttt{microJAX} offers a modular and differentiable framework for physically interpretable microlensing inference. With continued development, it has the potential to play a central role in analyzing next-generation microlensing datasets and advancing our understanding of planetary populations.

\section*{Acknowledgments}
We thank the anonymous referee for constructive feedback that helped improve the clarity of the paper. We are also grateful to Kohei Miyakawa, Yui Kasagi, Shotaro Tada, and Yamato Ureshino for productive discussions on differentiable programming. We also thank Kento Masuda and Ryou Ohsawa for valuable comments on this manuscript. This study was supported by the Japan Society for the Promotion of Science (JSPS) KAKENHI Grant Numbers 23K19079, 25KJ0428, and 25K17449. We acknowledge the use of ChatGPT (OpenAI) in improving the grammar and clarity of the manuscript.

\vspace{5mm}


\software{
\texttt{microJAX} \citep{microjax011}, \texttt{JAX} \citep{jax2018github}, \texttt{Numpyro} \citep{Bingham+2018, Phan+2019}, \texttt{VBBinaryLensing} \citep{Bozza2010, Bozza+2018}, \texttt{corner} \citep{corner} 
}

\appendix
\section{Integration Rules for Surface Brightness Profiles}
\label{appendix:integration}

\begin{figure}
    \centering
    \includegraphics[scale=0.17]{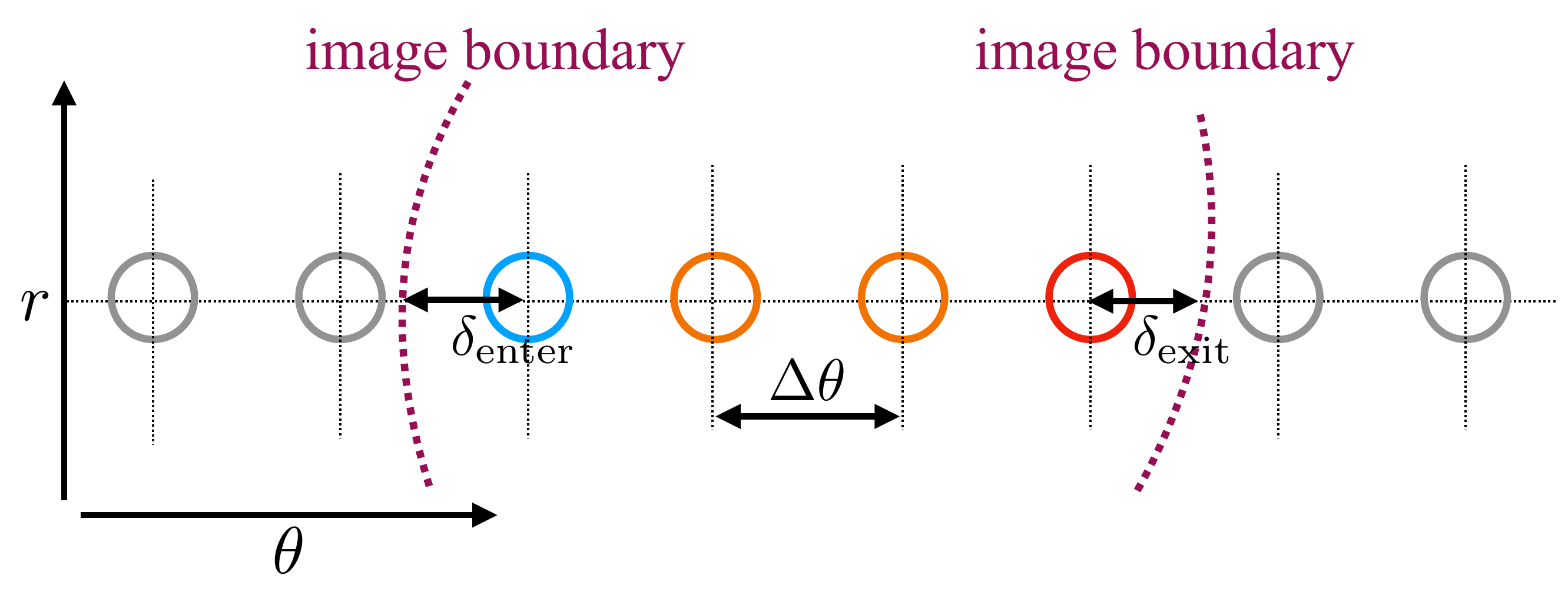}
    \caption{
    A schematic view of the integration method for a given $r$-row. The circles represent the grid points assigned for the inverse-ray shooting, and the purple lines represent the image boundary. The orange grid points are fully inside the image, the blue (entering) and red (exiting) ones are partially inside, and the gray ones are outside $(d>\rho)$. The boundary position of the image at each row, i.e., $d=\rho$, is estimated using cubic interpolation based on the distances of four nearest points to the boundary.
    }
    \label{fig:concept_ICRS}
\end{figure}

This appendix provides the full numerical formulation for computing the magnification in \texttt{microJAX}, using the inverse-ray shooting method on a fixed-resolution polar grid. Given the annular image regions defined in Section~\ref{sec:ICRS_rule}, we discretize each image into $N_r \times N_{\theta}$ grid points and compute their inverse-ray mapped positions in the source plane. The magnification is then obtained by numerically integrating the source brightness over these mapped positions, with quadrature rules adapted to the source profile.

\subsection{For Uniform Brightness Source Profile}
We describe here the method used to compute the magnification when the source has a uniform surface brightness profile. To accurately evaluate the contribution of each azimuthal segment, the algorithm examines local variations in the distance to the source center and uses sub-grid interpolation to resolve image boundaries with high precision. For each fixed radius $r_i$, we consider four consecutive azimuthal points, $(\theta_{\ell-1}, \theta_{\ell}, \theta_{\ell+1}, \theta_{\ell+2})$, along with their corresponding distances from the source center, $(d_{\ell-1}, d_{\ell}, d_{\ell+1}, d_{\ell+2})$. These values are used to construct a cubic Lagrange interpolant $d(\theta)$, which approximates the distance profile along the azimuthal direction. The boundary crossing angle $\theta_{\rm cross}$ is then estimated by solving $d(\theta_{\rm cross}) = \rho$, enabling precise identification of the image boundary. This interpolation enables accurate integration across segments that partially overlap the source disk, thereby reducing aliasing artifacts and enhancing the fidelity of the computed magnification. The use of cubic interpolation strikes a balance between numerical accuracy and computational efficiency, and is particularly effective for resolving sharp transitions near image edges.

Figure~\ref{fig:concept_ICRS} represents a schematic view of the grid allocation for a given $r$-row. To determine the contribution of each azimuthal segment, we classify adjacent angular pairs $(\theta_\ell, \theta_{\ell+1})$ based on whether their corresponding rays map inside or outside the source boundary. For each fixed radius $r_i$, we define an inclusion indicator
\begin{equation}
    \chi_\ell = \Theta(\rho - d_\ell),
\end{equation}
where $d_\ell = |\bm{w}_{i,\ell} - \bm{w}_{c}|$, and $\Theta$ is the Heaviside step function. Each segment is then categorized as follows:
\begin{itemize}[itemsep=0pt, left=0em, topsep=0pt, labelsep=0.5em]
    \item {\bf Fully enclosed} (\(\chi_\ell = \chi_{\ell+1} = 1\)): Both endpoints lie inside the source, and the segment contributes fully to the integral (orange-orange segments in Figure~\ref{fig:concept_ICRS}).
    \item {\bf Entering} (\(\chi_\ell = 0,\; \chi_{\ell+1} = 1\)): The segment crosses into the source; the contribution is determined via interpolation at the entry point (gray-blue segments).
    \item {\bf Exiting} (\(\chi_\ell = 1,\; \chi_{\ell+1} = 0\)): The segment crosses out of the source, treated analogously (red-gray segments).
\end{itemize}
For crossing segments, we estimate the boundary angle $\theta_{\rm cross}$ using cubic interpolation based on the distances of nearby points. The fractional coverage is then given by
\begin{equation}
    \delta_{\ell,{\rm enter}} = \frac{\theta_{\ell+1} - \theta_{\rm cross}}{\Delta\theta}, \quad
    \delta_{\ell,{\rm exit}} = \frac{\theta_{\rm cross} - \theta_\ell}{\Delta\theta},
\end{equation}
where $\Delta\theta$ is the azimuthal grid spacing. The total contribution from radius $r_i$ is computed by summing over azimuthal segments using a midpoint rule:
\begin{align}
    A_i = r_i \Delta\theta \sum_{\ell} \Bigl[ & \chi_\ell \chi_{\ell+1} 
    + (1 - \chi_\ell)\chi_{\ell+1} \left(\delta_{\ell,{\rm enter}} + \tfrac{1}{2}\right) \notag \\
    & + \chi_\ell (1 - \chi_{\ell+1}) \left(\delta_{\ell,{\rm exit}} + \tfrac{1}{2}\right) \Bigr].
\end{align}

Finally, the contributions from all radial rings are integrated using the trapezoidal rule to yield the area of the sector:
\begin{equation}
    A_{\rm sector} = \sum_{i=0}^{N_r - 1} A_i \Delta r,
\end{equation}
and the total magnification is given by summing over all sectors and normalizing by the area of the source disk:
\begin{equation}
    A = \frac{1}{\pi \rho^2} \sum_{\text{sectors}} A_{\rm sector}.
\end{equation}

\subsection{For Limb-darkened Source Profile}
We now extend the integration scheme to sources with a limb-darkened surface brightness profile. In this case, the magnification is computed by weighting each area contribution by the position-dependent source brightness $I(\tilde{r})$, where $\tilde{r} = d / \rho$ is the normalized radial distance from the source center. For the linear (first-order) limb-darkening law, the brightness is given by
\begin{equation}
    I(\tilde{r}) = I_0 \left[1 - u_1 \left(1 - \sqrt{1 - \tilde{r}^2} \right) \right],
\end{equation}
where $u_1$ is the limb-darkening coefficient, and $I_0 = 3 / \pi(3 - u_1)$ is a normalization factor that ensures the total flux over the source integrates to unity. The profile is defined for $0 \leq \tilde{r} \leq 1$.

As in the uniform-brightness case, each radial ring is divided into azimuthal segments. For each fixed radius $r_i$, we consider five consecutive angular grid points $(\theta_{\ell-2}, \theta_{\ell-1}, \theta_\ell, \theta_{\ell+1}, \theta_{\ell+2})$ and their corresponding distances $(d_{\ell-2}, d_{\ell-1}, d_\ell, d_{\ell+1}, d_{\ell+2})$ to the source center. Based on the inclusion flags $\chi_{\ell-1}, \chi_\ell, \chi_{\ell+1}$, each segment centered at $\theta_\ell$ (spanning $\theta_\ell \pm \Delta\theta/2$) is classified as follows:
\begin{itemize}[itemsep=2pt, left=0em, topsep=2pt, labelsep=0.5em]
    \item {\bf Fully enclosed} (\(\chi_{\ell-1} = \chi_\ell = \chi_{\ell+1} = 1\)):  
    The segment lies entirely within the source and contributes fully to the integral, weighted by \(I(\tilde{r}_\ell)\) at the midpoint.
    \item {\bf Entering} (\(\chi_{\ell-1} = 0,\; \chi_\ell = \chi_{\ell+1} = 1\)):  
    The image boundary lies between \(\theta_{\ell-1}\) and \(\theta_\ell\). The crossing angle \(\theta_{\rm cross}\), satisfying \(d(\theta_{\rm cross}) = \rho\), is estimated via cubic interpolation using four of the five neighboring points.  
    The fractional contribution is  $\delta_{\ell, {\rm enter}}=(\theta_{\ell+1} - \theta_{\rm cross})/\Delta\theta$.
    \item {\bf Exiting} (\(\chi_{\ell-1} = \chi_\ell = 1,\; \chi_{\ell+1} = 0\)):  
    The image boundary lies between \(\theta_\ell\) and \(\theta_{\ell+1}\). The crossing angle \(\theta_{\rm cross}\) is estimated as above, and the fractional contribution is $\delta_{\ell, {\rm exit}}=(\theta_{\rm cross} - \theta_{\ell})/\Delta\theta$.
\end{itemize}

A naive application of the trapezoidal rule to boundary-crossing segments leads to significant numerical errors, particularly near the limb of the source where the surface brightness gradient becomes steep. For the first-order limb-darkening law, the surface brightness in the image plane near the edge behaves approximately as $f(\bm{z}) \sim \sqrt{\bm{z} - \bm{z}_L}$, where $\bm{z}_L$ denotes a point on the image boundary. As revealed by a Taylor expansion, applying the trapezoidal rule to such functions introduces a leading-order integration error of $\mathcal{O}(\Delta\theta^{3/2})$, due to the divergence of the second derivative near the edge: $f''(\bm{z}) \propto (\bm{z} - \bm{z}_L)^{-3/2}$. This singular behavior invalidates the assumptions underlying the trapezoidal approximation, necessitating the use of more suitable integration rules tailored to such edge singularities. 

To reduce integration error while preserving the use of a fixed polar grid, we adopt a hybrid quadrature scheme inspired by \citet{Bennett+2010}. This method modifies the standard midpoint rule for segments that are only partially enclosed within the source disk. Specifically, each segment's contribution is scaled by a correction factor $f(\delta)$, which depends on the fractional coverage $\delta$ of the segment and is defined piecewise as:
\begin{eqnarray} 
f(\delta) = 
\begin{cases} 
    \left(\frac{1}{2} + \delta \right) b(\delta), & \text{if } \delta \geq \delta_c \\ 
    \frac{2}{3} \delta + \frac{1}{2}, & \text{if } \delta < \delta_c,
\end{cases} 
\end{eqnarray}
where $\delta_c$ is a tunable threshold chosen based on the empirical results of \citet{Bennett+2010}. This formulation offers a significant improvement in integration accuracy over the naive trapezoidal rule and also over the limiting cases $\delta_c = 0$ or $1$. The correction function
\begin{eqnarray}
    b(\delta) = \frac{2}{3} \sqrt{1 + \frac{1}{2\delta}}
\end{eqnarray}
is derived to approximate the integral of a square-root brightness profile near the limb of the source. For small $\delta$, the second branch of $f(\delta)$ ensures numerical stability and yields a second-order accurate approximation (see \citet{Bennett+2010} for details). Using this scheme, the area contribution for each radius $r_i$ is computed by summing over all azimuthal segments:
\begin{align} 
A_i = r_i \Delta\theta \sum_\ell I(\tilde{r}_\ell)\, \chi_\ell \Big[ 
    &\chi_{\ell-1} \chi_{\ell+1} 
    + (1 - \chi_{\ell-1}) \chi_{\ell+1} f(\delta_{\ell, \mathrm{enter}}) \notag \\
    &+ \chi_{\ell-1} (1 - \chi_{\ell+1}) f(\delta_{\ell, \mathrm{exit}})
\Big], 
\end{align}
where $I(\tilde{r}_\ell)$ is the limb-darkened surface brightness at the segment center. The radial integration is then performed using the trapezoidal rule: $A_{\mathrm{sector}} = \sum_{i=0}^{N_r-1} A_i \Delta r$, and the final magnification is obtained by normalizing the total lensed flux by the intrinsic source flux: $A = \sum_{\mathrm{sector}} A_{\mathrm{sector}}/\rho^2$, assuming the source brightness profile is already normalized to unit total flux.

\section{Marginalizing Linear Flux Parameters With Non-Linear Parameter Inference\label{appendix:marginalize}}

\subsection{Bayesian Formulation}
Following \citet{Hogg+2021} and \citet{Bartolic2023}, consider a dataset $\bm{y} = \{y_i\}$ with covariance matrix $\bm{C}$, and a model that includes nonlinear parameters $\bm{\theta}$ and linear parameters $\bm{\beta}$. We are interested in evaluating the likelihood: $p(\bm{y}|\bm{\theta}, \bm{\beta})$. However, if we are not interested in the linear parameters themselves, the computational cost of inference can be reduced if the following marginalization is tractable:
\begin{eqnarray}
    p(\bm{y}|\bm{\theta}) = \int p(\bm{y}|\bm{\theta}, \bm{\beta})\, p(\bm{\beta}|\bm{\theta})\, d\bm{\beta}.
\end{eqnarray}
This integral requires a prior distribution $p(\bm{\beta}|\bm{\theta})$ that may depend on the nonlinear parameters. Alternatively, one can use Bayes' theorem to rewrite the integral in the following form:
\begin{eqnarray}
    p(\bm{y}|\bm{\theta}, \bm{\beta})\, p(\bm{\beta}|\bm{\theta}) &=& p(\bm{\beta}|\bm{y}, \bm{\theta})\, p(\bm{y}|\bm{\theta}) \\
    \Rightarrow p(\bm{y}|\bm{\theta}) &=& \frac{p(\bm{y}|\bm{\theta}, \bm{\beta})\, p(\bm{\beta}|\bm{\theta})}{p(\bm{\beta}|\bm{y}, \bm{\theta})}.
\end{eqnarray}
This expression enables the evaluation of the marginalized likelihood without explicitly performing the integral. If both the prior and the likelihood are multivariate Gaussian distributions, the expressions become:
\begin{eqnarray}
    p(\bm{y}|\bm{\theta}, \bm{\beta}) &&= \mathcal{N}(\bm{y}|\bm{M}\bm{\beta}, \bm{C}) \notag\\ &&\propto \exp\left( -\frac{1}{2} (\bm{y} - \bm{M}\bm{\beta})^\top \bm{C}^{-1} (\bm{y} - \bm{M}\bm{\beta}) \right) \notag\\
    p(\bm{\beta}|\bm{\theta}) &&= \mathcal{N}(\bm{\beta}|\bm{\mu}, \bm{\Lambda}) \notag \\ 
    && \propto \exp\left( -\frac{1}{2} (\bm{\beta} - \bm{\mu})^\top \bm{\Lambda}^{-1} (\bm{\beta} - \bm{\mu}) \right)\notag
\end{eqnarray}
\begin{align}
    \Rightarrow &-\ln \mathcal{N}(\bm{y}|\bm{M}\bm{\beta}, \bm{C}) \, \mathcal{N}(\bm{\beta}|\bm{\mu}, \bm{\Lambda}) \notag \\ 
    &\propto  (\bm{y} - \bm{M}\bm{\beta})^\top \bm{C}^{-1} (\bm{y} - \bm{M}\bm{\beta}) + (\bm{\beta} - \bm{\mu})^\top \bm{\Lambda}^{-1} (\bm{\beta} - \bm{\mu}). 
\end{align}
The marginalized likelihood $p(\bm{y}|\bm{\theta})$ can then be evaluated as:
\begin{align}
    \mathcal{N}(\bm{y}|\bm{M}\bm{\beta}, \bm{C})\, \mathcal{N}(\bm{\beta}|\bm{\mu}, \bm{\Lambda})
    = \mathcal{N}(\bm{\beta}|\bm{a}, \bm{A})\, \mathcal{N}(\bm{y}|\bm{b}, \bm{B}),
\end{align}
where
\begin{eqnarray}
    \bm{A}^{-1} &=& \bm{\Lambda}^{-1} + \bm{M}^\top \bm{C}^{-1} \bm{M}, \notag \\
    \bm{a} &=& \bm{A} (\bm{\Lambda}^{-1} \bm{\mu} + \bm{M}^\top \bm{C}^{-1} \bm{y}), \notag \\
    \bm{B} &=& \bm{C} + \bm{M} \bm{\Lambda} \bm{M}^\top, \notag \\
    \bm{b} &=& \bm{M} \bm{\mu}. \notag
\end{eqnarray}
Thus, since the term involving the linear parameter $\bm{\beta}$ can be integrated out, we obtain:
\begin{align}
p(\bm{y}|\bm{\theta}) = \int \mathcal{N}(\bm{\beta}|\bm{a}, \bm{A})\, \mathcal{N}(\bm{y}|\bm{b}, \bm{B})\, d\bm{\beta} = \mathcal{N}(\bm{y}|\bm{b}, \bm{B}).
\end{align}

\subsection{Application to Microlensing Modeling}
To model the observed fluxes in a gravitational microlensing event, we introduce a hierarchical framework in which each dataset $j$ is described by a set of linear flux parameters $\bm{\beta}_j$ and shared non-linear parameters $\bm{\theta}$ that characterize the lensing model. Assuming independent Gaussian noise for each dataset, the joint likelihood can be written as
\begin{align}
    p(\bm{D|\bm{\beta, \bm{\theta}}}) = \prod^J_{j=1} p(\bm{F}_j | \bm{\beta}_j, \bm{\theta})
\end{align}
where $J$ and $j$ refer to the number of datasets and their index, respectively. When we assume Gaussian priors for linear parameters of flux parameters $\bm{\beta}$, we can marginalize the probabilities of linear parameters for the non-linear parameter inference \citep{Hogg+2020}. In this formulation, the design matrix $\bm{M}_j$ depends only on the non-linear parameters $\bm{\theta}$, and relates them to the expected contribution of the linear flux parameters.
Using the formula in the previous section, the negative log-likelihood (NLL) for non-linear parameters can be written as 
\begin{eqnarray}\label{eq:NLL}
    \mathrm{NLL}(\bm{\theta}) &=& \frac{1}{2}(\bm{F}_j - \bm{M}_j\bm{\mu})^\mathsf{T} (\bm{C}_j + \bm{M}_j\bm{\Lambda}\bm{M}^\mathsf{T}_j)^{-1} (\bm{F}_j - \bm{M}_j\bm{\mu}) \nonumber \\&&+ \frac{1}{2} \ln | \bm{C}_j + \bm{M}_j\bm{\Lambda}\bm{M}^\mathsf{T}_j|
\end{eqnarray}
where 
\begin{align}
    &(\bm{C}_j + \bm{M}_j\bm{\Lambda}\bm{M}^\mathsf{T}_j)^{-1} \nonumber\\ 
    &= \bm{C}_j^{-1} - \bm{C}_j^{-1}\bm{M}_j(\bm{\Lambda}^{-1} + \bm{M}^\mathsf{T}_j\bm{C}^{-1}_j\bm{M}_j)^{-1} \bm{M}_j \bm{C}_j^{-1} \\
    &\ln | \bm{C}_j + \bm{M}_j\bm{\Lambda}\bm{M}^\mathsf{T}_j|\nonumber \\ 
    &= \ln{|\bm{C}_j|} + \ln{|\bm{\Lambda}|} + \ln{|\bm{\Lambda}^{-1} + \bm{M}^\mathsf{T}_j\bm{C}^{-1}_j\bm{M}_j|},
\end{align}
which are derived from the matrix inversion lemma, also known as the Woodbury matrix identity.

Regarding a single-source microlensing model:
\begin{align}
    \{\bm{F}\}_j = \left[\bm{A}(\bm{t}_{j}; \bm{\theta}) - 1\right] F_{S,j} + F_{\mathrm{base},j},
\end{align}
marginalizing over the linear parameters, we define
\begin{align}
    \bm{M}_j &= \begin{pmatrix}
       A(t_{j,1}; \bm{\theta}) - 1 & 1 \\
       \vdots \\
       A(t_{j,n_j}; \bm{\theta}) - 1 & 1
    \end{pmatrix}\in \mathbb{R}^{n_j \times 2}, \\
    \bm{\mu}_j &= (
        F_{S,j} , F_{\mathrm{base},j})^\top, \\
    \bm{\Lambda} &= 
        (\sigma^2_{F_S,j}, \sigma^2_{F_\mathrm{base},j})^\top,
\end{align}
and by substituting these expressions, it becomes possible to estimate the microlensing parameters without directly estimating the linear parameters $F_S$ and $F_{\mathrm{base}}$. In general, since we can assume that $\bm{\mu}=\bm{0}$, $\bm{\Lambda}$ and $\bm{C}_j$ are diagonal, the evaluation of Equation \eqref{eq:NLL} is not expensive to compute. 

If you want to obtain the posterior distribution of the linear parameters, you can sample from the following distribution:
\begin{eqnarray} 
    p(\bm{\beta} \mid \bm{y}, \bm{\theta}) &=& \mathcal{N}(\bm{\beta} \mid \bm{a}(\bm{\theta}), \bm{A}(\bm{\theta})) \\
    \bm{A}(\bm{\theta})^{-1} &=& \bm{\Lambda}^{-1} + \bm{M}(\bm{\theta})^\top \bm{C}^{-1} \bm{M}(\bm{\theta}) \\ 
    \bm{a}(\bm{\theta}) &=& \bm{A}(\bm{\theta}) \left[ \bm{\Lambda}^{-1} \bm{\mu} + \bm{M}(\bm{\theta})^\top \bm{C}^{-1} \bm{y} \right] 
\end{eqnarray}



\bibliography{0_main}{}
\bibliographystyle{aasjournal}



\end{document}